\documentclass[superscriptaddress, nofootinbib]{revtex4-2}
\usepackage[utf8]{inputenc}
\usepackage{appendix,typearea,amsmath,color,url,epsfig,graphicx,amsthm,ulem,xcolor}

\usepackage{graphicx} %
\usepackage{amsthm}
\usepackage{amsmath}
\usepackage{amssymb}
\usepackage{amsthm}
\theoremstyle{plain}

\usepackage{xcolor}
\usepackage{graphicx}
\usepackage{multirow}
\usepackage{subcaption}

\theoremstyle{definition}
\newtheorem{ex}{Example}

\newtheorem{rem}{Remark}

\typearea{14}
\begin{document}
\title[Mutual information via quantum kernel method]{Estimation of mutual information via quantum kernel method}
\author{Yota~Maeda}
\email{yota.maeda@sony.com}
\affiliation{Advanced Research Laboratory, Technology Infrastructure Center, Technology Platform, Sony Group Corporation, 1-7-1 Konan, Minato-ku, Tokyo, 108-0075, Japan}
\affiliation{Quantum Computing Center, Keio University, Hiyoshi 3-14-1, Kohoku-ku, Yokohama 223-8522, Japan}

\author{Hideaki~Kawaguchi}
\email{hikawaguchi@keio.jp}
\affiliation{Quantum Computing Center, Keio University, Hiyoshi 3-14-1, Kohoku-ku, Yokohama 223-8522, Japan}
\affiliation{Human Biology-Microbiome-Quantum Research Center (WPI-Bio2Q), Keio University, Mita 2-15-45, Minato-ku, Tokyo 108-8345, Japan}

\author{Hiroyuki~Tezuka}
\email{hiroyuki.tezuka@sony.com}
\affiliation{Advanced Research Laboratory, Technology Infrastructure Center, Technology Platform, Sony Group Corporation, 1-7-1 Konan, Minato-ku, Tokyo, 108-0075, Japan}
\affiliation{Quantum Computing Center, Keio University, Hiyoshi 3-14-1, Kohoku-ku, Yokohama 223-8522, Japan}

\begin{abstract}
Recently, the importance of analysing data and collecting valuable insight efficiently has been increasing in various fields.
Estimating mutual information~(MI) plays a critical role to investigate the relationship among multiple random variables with a nonlinear correlation.
Particularly, the task to determine whether they are independent or not is called the independence test, whose core subroutine is estimating MI from given data.
It is a fundamental tool in statistics and data analysis that can be applied in a wide range of application such as hypothesis testing, causal discovery and more. 
In this paper, we propose a method for estimating mutual information using the quantum kernel.
We investigate the performance under various problem settings, such as different sample size or the shape of the probability distribution.
As a result, the quantum kernel method showed higher performance than the classical one under the situation that the number of samples is small, the variance is large or the variables posses highly non-linear relationships.
We discuss this behavior in terms of the central limit theorem and the structure of the corresponding quantum reproducing kernel Hilbert space.
\end{abstract}

\maketitle

\section{Introduction}
\label{section:intro}
In classical statistics, the correlation coefficient is an important quantity for examining the correlations between variables. 
However, this indicator can only detect linear relationships.
For the nonlinear expression, it needs to be generalized as the \textit{mutual information} $I(x,y)$.
The mutual information expresses how two random variables $x,y$ relate based on the distance between the random distributions, for instance, the Kullback-Leibler divergence or Pearson divergence.
A method of determining the independence between random variables based on this mutual information is called an independence test, and the mutual information used in this process is called the \textit{mutual information criterion}.
The independence test is known to have a wide range of applications. 
One of the most significant application is the causal  discovery algorithm.
It reveals causal relationships that cannot be distinguished by simply looking at the dataset.
That task often offers a suggestive interpretation from a complicated dataset, and is used as an effective means in data science these days. 
In Direct LiNGAM~\cite{shimizu2011directlingam}, a typical causal discovery algorithm, one checks the independence of variables and residuals by estimating the mutual information after performing linear regression, and the causal order is determined according to the Darmois-Skitovich theorem~\cite{2810c4a1-6916-356b-a07a-96c671cc4dad, skitovitch1953property}.
There are several methods for estimating the amount of mutual information, and the most common way is based on the kernel method.

In this paper, we perform two types of kernel-based estimation of the mutual information, including classical and quantum kernel methods.
As a result, the quantum method shows comparable performance to the classical one for the linear model, and shows superior performance under certain conditions, such that the probability distribution has large variances.
We also discuss whether the circuit used in~\cite{kawaguchi2023application} to estimate the Hilbert-Shimidt norm of a normalized cross-covariance operator is characteristic. 
This refers to the functional properties of the quantum kernel resulting from a particular quantum circuit.

The mutual information criterion can be formulated in several ways, and we simply represent it as $I(x,y)$
in this paper.
The quantity $I(x,y)$ must have the following properties generally:
\begin{enumerate}
    \item (Symmetry) $I(x,y)=I(y,x)$.
    \item (Positivity) $I(x,y)\geq 0$.
    \item (Independent-ness) $I(x,y)=0$ if and only if $x$ and $y$ are independent.
\end{enumerate}
As the quantities with these properties, we introduce two types of criteria. %
The well-known formulation of the mutual information~(MI)~\cite{shannon1948mathematical, cover1999elements} is
\begin{align}
    \label{eq:common_mi}
    MI(X,Y):=\int_{\mathcal{X}}\int_{\mathcal{Y}}p_{XY}(x,y)\log\frac{p_{XY}(x,y)}{p_X(x)p_Y(y)}dxdy,
\end{align}
where $X$ and $Y$ are the random variables sampled from the probability distribution $\mathcal{X}$ and $\mathcal{Y}$, respectively.
This quantity is based on the Kullback-Leibler divergence.
It has a sharp shape at the origin since it contains a logarithmic function, and thus is sensitive to small changes in the probability density functions~\cite{sugiyama2012machine}.
It is problematic for some tasks such as anomaly detection  since the estimation of the mutual information from a sample is easily affected by anomalies~\cite{basu1998robust, sugiyama2012density}.

To solve this problem, in this paper, we also discuss the squared-loss mutual information (SMI) that was first introduced in~\cite{suzuki2009mutual}:
\begin{align}
\label{eq:smi}
SMI(X,Y):=\int_{\mathcal{X}}\int_{\mathcal{Y}}p_X(x)p_Y(y)\left(\frac{p_{XY}(x,y)}{p_X(x)p_Y(y)}-1\right)^2dxdy.
\end{align}
As noted in~\cite{sugiyama2012machine}, this quantity has several superior points, including the robustness against outliers~\cite{basu1998robust, sugiyama2012density}.
Since the ratio of probability density functions tends to be a steep function, it is not always easy to accurately estimate the squared loss mutual information from a small number of samples~\cite{sakai2014computationally}.

Estimating mutual information is important in a wide range of areas such as feature selection \cite{fang2015feature}, clustering \cite{slonim2005information}, gene networking \cite{jiang2022gene}, anomaly detection \cite{kopylova2008mutual} and causal discovery, which we will discuss in detail in this paper.
However, the accurate estimation of the mutual information is known to be difficult in practice~\cite{kraskov2004estimating}.
When we estimate MI, we can use the theory of entropy~\cite{cover1999elements}.
The most common method via estimation of the entropy is the KSG estimator~\cite{kraskov2004estimating}, which is a variant of the kNN estimator using max-norm distance.
However, estimating entropy heavily relies on the classical statistical methods, hence have some problems \cite{gao2015efficient}. 
First, it needs a sufficiently large amount of data compared to kernel methods;
\cite[Theorems 2, 3]{gao2015efficient} showed that when estimating nonparametric mutual information accurately via using the classical kNN estimator, there are cases where exponentially many samples are necessary.
Moreover, reliable estimation can be difficult, especially for high-dimensional data or data with rare events.
Second, there is uncertainty in estimation; entropy estimation is a statistical method based on sample data. 
Therefore, the estimates themselves are also subject to uncertainty. 
Small sample sizes can lead to large estimation errors.
More advanced information-theoretic methods could capture nonlinear relationships better.
As we will see in Sec. \ref{section:mi}, we estimate these criteria by the kernel methods.

Estimating kernel methods using classical computers often requires large amounts of data. Furthermore, this method often uses a Gaussian kernel, for which the data must have a shape similar to a normal distribution. However, otherwise, it is often difficult to determine the independence of the input data using the classical kernel method. On the other hand, these difficulties can sometimes be solved by using quantum kernel methods, which are kernel methods based on quantum computers.
In this paper, we discuss how to choose the kernel and the mutual information criteria in terms of the number of samples, the model complexity, data embedding, and properties of classical or quantum kernels.

This paper is organized as follows.
In Sec. \ref{section:mi}, we introduce mutual information criteria. 
Next, we review the kernel method in Sec. \ref{section:kernel}. 
In particular, we discuss what kind of quantum circuit to choose when using the quantum kernel method.
In our work, from the viewpoint of computational theory, we will use a quantum circuit based on the IQP circuit, which is considered difficult to efficiently simulate the sampled discrete probability distribution by classical computers.
We further explain how to estimate them from samples using the kernel methods. 
The novel part is the discussion of the structure of the reproducing kernel Hilbert space (RKHS) determined from the quantum kernel method for the estimation of SMI.
Based on that, we experiment to what extent we can distinguish between random and dependent variables with linear and nonlinear relational expressions in Sec. \ref{section:experiments}.
Then, we analyze them in terms of classes of kernels and concrete forms of series expansion in Sec. \ref{section:analysis}.
Finally, we refer to applications to the causal discovery in Sec.  \ref{section:application} and discuss future work in Sec.  \ref{section:future_work}.

\section{Mutual information criteria}
\label{section:mi}
In this section, we explain two mutual information criteria; the naive formulation 
 (Subsec. \ref{subsection:mi}) and the squared-loss one (Subsec. \ref{subsection:smi}).

\subsection{Mutual information}
\label{subsection:mi}
The most common measure of the difference between the two probability distributions $p$ and $q$ is the Kullback-Leibler divergence, described below.
\[\int_{-\infty}^{\infty}p(x)\log\frac{q(x)}{p(x)}.\]
Based on this, the mutual information can be defined as (\ref{eq:common_mi}).
A priori, this quantity is described from the entropy $H(\cdot)$ of the random variables~\cite{cover1999elements} as
\[MI(X,Y)=H(X)+H(Y)-H(X:Y).\]
To estimate the mutual information from given data, one can either use the above method based on entropy estimation or the kernel method. 
In the case of the entropy-based method, the probability distribution must be estimated directly based on the statistical properties of the data. 
There are four drawbacks to this method as follows:
\begin{enumerate}
    \item (Data volume requirements) Sufficiently large amounts of datasets are needed to obtain more accurate estimates. A small amount of dataset may lead to inaccurate estimation of probability distributions, and thus to less accurate entropy estimation.
    \item (Influence of missing values and noise) If a dataset contains missing values or noise, the estimated probability distribution may be skewed. This can reduce the accuracy of entropy estimation. In particular, when there are many missing values or noise, it is difficult to accurately estimate the probability distribution.
    \item (The curse of dimensionality) Entropy estimation tends to become more difficult as the dimensionality of the dataset increases. As the dimensionality of the dataset increases, the accuracy of the estimated probability distribution decreases and entropy estimation may become less reliable. This phenomenon is known as the ``curse of dimensionality''.
    \item (Lack of prior knowledge) Entropy estimation requires sufficient prior knowledge of the shape and parameters of the probability distribution. However, in real problems, the dataset may follow an unknown probability distribution. In such cases, it becomes difficult to select appropriate probability distribution assumptions and parameter estimation methods.
\end{enumerate}
Bach-Jordan proposed a method to estimate MI \eqref{eq:common_mi} using the kernel methods~\cite{bach2002kernel}, and still the kernel methods have been actively studied to solve these problems.
\begin{rem}
    Bach-Jordan pointed out that the Gaussian kernel is powerful, particularly in a situation where the variance of the estimated probability distribution is small.
    However, there exist various situations not applicable to that condition.
    Therefore, in this paper, we aim to find the effective kernel function even for situations far from that condition.
\end{rem}
We propose a method to determine the independence between two random variables by the quantum kernel method, hence we review~\cite[Subsec. 3.2]{bach2002kernel} here; see also~\cite[Sec. 3]{shimizu2011directlingam}.
For two given random variables $x_1, x_2 \in \mathbb{R}^n$, we obtain the corresponding Gram matrices $K_1$ and $K_2$.
Note that according to the choice of the kernel $k$, the $(i,j)$-entry of the matrix $K_{\ell}$ ($\ell=1,2$) can be described as 
\[k(x_{\ell}^{(i)}, x_{\ell}^{(j)}).\]
Let us define two matrices 
\begin{align*}
    \mathcal{K}_K&:=\begin{pmatrix}
\left(K_1 + \frac{n\kappa}{2}I\right)^2 & K_1K_2 \\
K_2K_1 & \left(K_2 + \frac{n\kappa}{2}I\right)^2 \\
\end{pmatrix}\\
    \mathcal{D}_K&:=\begin{pmatrix}
\left(K_1 + \frac{n\kappa}{2}I\right)^2 & 0 \\
0 & \left(K_2 + \frac{n\kappa}{2}I\right)^2 \\
\end{pmatrix}
\end{align*}
for a small hyperparameter $\kappa>0$, which plays a role as a regularization.
These are derived from the generalized eigenvalue problem inspired by the kernel-based principal component analysis~\cite{scholkopf1998nonlinear} and studied in canonical correlation analysis.
Here, it is known that the quantity
\[\widehat{MI}(x_1,x_2):= -\frac{1}{2}\log\frac{\mathrm{det}\mathcal{K}_k}{\mathrm{det}\mathcal{D}_k}\]
is a good estimator of the mutual information $MI(X,Y)$ proved in~\cite[Appendix]{bach2002kernel}.
Optimally, when using the Gaussian kernel as $k$, one should take its width $\sigma\leq 1$ as described in \cite{bach2002kernel}.
This is supported by experimental results in \cite{shimizu2011directlingam, bach2002kernel} and theoretical considerations~\cite[Subsec. 4.5]{bach2002kernel}.

\subsection{Squared-loss mutual information}
\label{subsection:smi}
In this section, we review another method to estimate SMI using the Hilbert-Shmidt norms of the normalized covariance operator~\cite{gretton2005measuring}.
Fukumizu et al.~\cite{fukumizu2007kernel} generalized this work in terms of kernel methods and proposed a method to estimate SMI, independently of the concrete form of kernels.
It is known that the conventional method for estimating mutual information using kernel methods  (including one in Subsec. \ref{subsection:smi}) is highly dependent on the shape of the kernel. 
They showed that their method is independent of the kernel under the condition that the kernel is \textit{characteristic}.
This means that the estimator coincides with SMI in the limit of a large number of samples.

Let $k$ be a kernel on a set of data $\mathcal{X}$.
Let us denote by $\mathcal{H}_k$ be the corresponding RKHS.
Then, by the Riesz representation theorem~\cite{conway2019course}, we can find a \textit{mean element} $m_k(P)$ satisfying
\begin{align}
\label{eq:mean_element}
    \langle f, m_k(P)\rangle=E[f(P)]
\end{align}
for any probability measure $P$ on $\mathcal{X}$ and $f\in\mathcal{H}_k$.

\begin{ex}[{\cite{fukumizu2010introduction}}]
    We compute an explicit form of a mean element for the case of the linear kernel $k(x,y)=(xy+c)^d$.
    In this situation, the RKHS $\mathcal{H}_k$ corresponds with the vector space consisting of polynomials with degrees less than $d+1$.
    Let $X$ be a random variable on $\mathbb{R}$.
    We denote by $\mu_r:=E[X^r]$ the moment of degree $r$ for $0\leq r\leq d$.
    Then, the mean element has an information of the moments of $X$, that is, 
    \[\mu_r=\langle x^r, m_k(X) \rangle\]
    for $f(x)=x^r$ in (\ref{eq:mean_element}).
\end{ex}

Let $\mathcal{P}$ be the set of probability measures on $\mathcal{X}$.
We call the kernel $k$ \textit{characteristic} if the following map is injective:
\begin{align*}
    \mathcal{P}&\to\mathcal{H}_k\\
    P&\mapsto m_k(P). 
\end{align*}
In other words, one can recover the original probability distribution from its mean element if the kernel is characteristic.
For instance, on the one hand, the Gaussian and Laplace kernels are known to be universal~\cite{steinwart2001influence}, and hence they are characteristic.
On the other hand, because of the scarce structure of RKHS associated with polynomial kernels, they are not characteristic.
This is due to the fact that the moments corresponding to the polynomial kernel do not have sufficient information on higher-order terms.
Roughly speaking, a kernel is universal if the corresponding RKHS is dense in the function space over $\mathcal{X}$, which can often be seen from the coefficients of the Taylor expansion of the kernel~\cite[Corolarries 10, 11]{steinwart2001influence}.

Next, we recall the (normalized) cross-covariance operator.
For the detailed discussion below, see~\cite[Sec. 2]{fukumizu2007kernel}.
Roughly speaking, as a generalization of the fact that the covariance describes a linear correlation of a given data, it represents a (non-)linear correlation between probability distributions with values in the RKHS.
Let $\mathcal{H}_{k_1}$ and $\mathcal{H}_{k_2}$ be RKHSs.
By the above discussion, there are specific elements $m_1$ and $m_2$.
Combined with the Cauchy-Schwarz inequality, we also have a mean element $m_{12}$ on $\mathcal{H}_1\otimes\mathcal{H}_2$.
Here, the element $m_{12}-m_1\otimes m_2$ is called the covariance in $\mathcal{H}_1\otimes\mathcal{H}_2$ and defines a linear map between $\mathcal{H}_1$ and $\mathcal{H}_2$.
The corresponding operator $\Sigma_{12}$ is called \textit{cross-covariance operator}, and we obtain the correlation coefficient by normalizing the covariance.
Even for the nonlinear relationship between the random variables, we can define the normalized cross-covariance operator (NOCCO) $V_{12}$ as the unique element satisfying the following equation:
\[\Sigma_{12} = \Sigma_{11}^{1/2} V_{12} \Sigma_{22}^{1/2}.\]
It is known that the Hilbert-Shmidt norm of NOCCO converges to SMI under some assumptions on the kernel and data~\cite[Theorems 4, 5]{fukumizu2007kernel}.
To state this explicitly, we describe the procedure for estimating  SMI by taking matrices based on the empirical distribution of the kernel values obtained from the dataset set.
Let $G_{\ell}$ ($\ell=1,2$) be the centered Gram matrices, where each entry is given by the difference between the kernel value and the empirical mean element.
Now, taking $R_{\ell}:=G_{\ell}(G_{\ell}+n\epsilon_nI_n)^{-1}$ for a decreasing hyperparameter sequence $\epsilon_n$ with $\epsilon_n\to 0$ and $\epsilon_n^3 n\to\infty$, we define
\begin{equation}
  \widehat{SMI}_n:=||V_{12}||^2_{HS}=\mathrm{Tr}(R_1R_2).
  \label{eq:smi_estimated}
\end{equation}
Note that this corresponds to $\widehat{I}^{NOCCO}_n$ described in~\cite{fukumizu2007kernel}.
Then, under the assumption that $k$ is characteristic, the above quantity $\widehat{SMI}_n$ converges to $SMI$ when $n\to\infty$, independent of the choice of $k$.

\section{Kernel methods}
\label{section:kernel}
\subsection{General theory}
We use kernel methods in this paper to estimate mutual information introduced in Sec. \ref{section:mi}. There are kernel methods that use classical or quantum computers.
Before discussing the quantum kernel method, let us introduce the general theory of the kernel method, including the ``classical" kernel.
Kernel methods are a class of algorithms used in machine learning and pattern recognition tasks. 
They are particularly effective for solving problems involving nonlinear relationships in the data.
For more details; see~\cite{campbell2002kernel, rasmussen2006gaussian, shawe2004kernel, scholkopf2001generalized}.
Let $\mathcal{X}$ be a set, commonly consisting of input data.
We call a function 
\[k:\mathcal{X}\times\mathcal{X}\to\mathbb{R}\]
a \textit{positive definite kernel} if it satisfies symmetry and (semi-)positive definiteness.
In this paper, we call it simply \textit{kernel}.
For such a function and subset of data $\{(x_i,x_j)\}_{1\leq i,j\leq n}\subset\mathcal{X}\times\mathcal{X}$, the matrix $K_k$, whose $(i,j)$-entry $(k(x_i,x_j))_{1\leq i,j\leq n}$, called the \textit{Gram matrix} is symmetric and semi-positive definite.
Then, by the Moore Aronszajn theorem~\cite{aronszajn1950theory, berlinet2011reproducing}, there exists a unique RKHS $\mathcal{H}$ with $k$ as the reproducing kernel.
The corresponding map called the \textit{feature map}
\begin{align}
\label{mor:feature_map}
    \phi:\mathcal{X} &\to \mathcal{H}_k\\
    x&\mapsto k(x, \cdot)
\end{align}
introduces, roughly speaking, a method for linear separation of a dataset to apply multivariate analysis such as principal component analysis~(PCA) or support vector machine~(SVM), which we will explain below, in the high-dimensional space $\mathcal{H}_k$.

One key advantage of kernel methods is that they can treat complex relationships between data points to be captured without explicitly defining the mapping function. 
This makes them well-suited for dealing with high-dimensional data or situations where the data resides in a space where direct calculations are difficult or impossible. 
Additionally, kernel methods have solid theoretical foundations and offer good generalization capabilities.
A kernel represents the similarity of the given dataset, and as an important property, the representer theorem, which holds under general conditions, is widely known~\cite{smola1998connection, scholkopf2002learning}.
By this theorem, the function that minimizes the loss function (with regularization) can be obtained by the kernel. 
This is called the \textit{kernel trick}, which allows us to analyze data only with the value of $k(x_i,x_j)$, and without directly calculating the feature map \eqref{mor:feature_map}.

Kernel methods operate on the assumption that the data points lie in a linearly separable feature space, even if they are not linearly separable in the original input space. By using a suitable kernel function, the data can be implicitly mapped into a higher-dimensional space where it becomes linearly separable. 
In this transformed space, linear algorithms, such as SVM, can be employed to find the optimal decision boundaries or classification rules~\cite{bottou2007large, campbell2002kernel, smola2000advances}.
When we apply kernel methods, we need to design a kernel function appropriately, since different kernels have different properties, and it directly affects the model's performance.
The most commonly used kernel is the Gaussian radial basis function (RBF) kernel, also known as the Gaussian kernel:
\begin{align}
    k(x,y):=\exp\left(-\frac{||x-y||^2}{2\sigma^2}\right),
    \label{eq:rbf_kernel}
\end{align}
where $x$ and $y$ are data points, and $\sigma$ is the variance.
It calculates the similarity between two points based on their Euclidean distance in the input space. Other popular kernel functions include the polynomial kernel, sigmoid kernel, and Laplacian kernel, each suited for specific types of data and problem domains~\cite{campbell2002kernel}.

\subsection{Quantum kernel method}
\begin{figure}[htb]
    \centering
    \includegraphics[width=5cm]{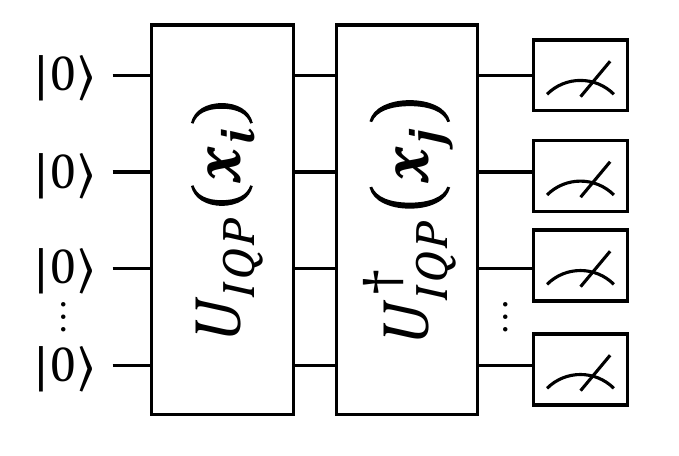} 
    \caption{The quantum circuit used in [2] for calculating inner product.}
    \label{figure:iqp}   
\end{figure}

In this subsection, let us briefly introduce the notion of quantum kernel methods.
For a general theory and its applications, see~\cite{chatterjee2016generalized, havlivcek2019supervised, schuld2019quantum, schuld2021machine}.
The quantum kernel is computed from two quantum states in which data is embedded.
Given a dataset $\mathcal{X}$, let us encode them into quantum states using quantum circuits.
If we write this correspondence $\rho$, we obtain a quantum state $\rho(\mathbf{x})$ for a data $\mathbf{x}\in\mathcal{X}$, which is an analog to the classical feature mapping.
We compute the quantum kernel as $Q(\mathbf{x}, \mathbf{x'}) = h(\rho(\mathbf{x}), \rho(\mathbf{x'}))$, where $h$ is a real-valued function that takes two density operators.
If we can compute the inner product between quantum states on a quantum computer, we can enjoy the merit of the kernel method.
This is called the \textit{quantum kernel method}.
Quantum kernel methods have the advantage that they can be  efficiently calculate the inner-product even with high dimensional feature space, as long as the unitary circuit for computing the inner product can be implemented (e.g. FIG.\ref{figure:iqp}).
To explore its advantage, the IQP circuit~\cite{bremner2016average} is used for a data embedding circuit since the output distribution generated by this kind of circuit cannot be efficiently sampled with classical computers under a reasonable assumption of complexity theory~\cite{bremner2011classical}.
For instance, \cite{havlivcek2019supervised} introduces the IQP circuit composed of ZZFeaturemap.
In addition to that, we embed input data with activation function in the same way as~\cite{kawaguchi2023application}.
Although there are few works on quantum kernel methods with activation functions, it largely affects, and could enhance, the machine learning performance~\cite{goto2021universal}.
In this paper, we investigate the performance of this type of quantum kernel for the independence test with comparing to the (classical) RBF kernel.
For the concrete form of the IQP circuit and an activation function, see~\cite{kawaguchi2023application}.

\section{Experiments}
\label{section:experiments}
\subsection{Experimental setting}
\begin{figure}[htb]
    \centering
    \includegraphics[width=5cm]{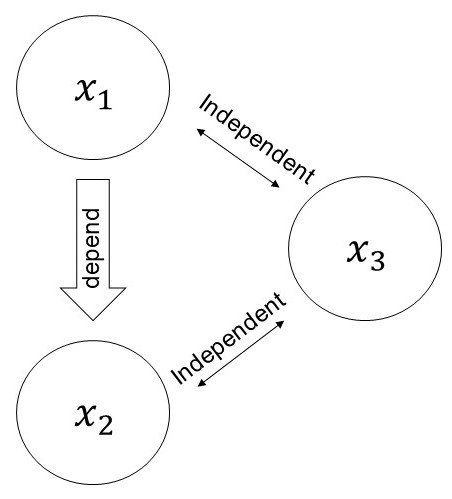} 
    \caption{The actual relationship between the variables.}
    \label{figure:experiment_setting}   
\end{figure}

Based on the mutual information (MI, SMI), we performed the experiments of independence tests among several types of probability distribution from a given data set using the kernel method.
The goal of this task is to determine the relationship among three variables shown in FIG. \ref{figure:experiment_setting}, called a causal model.
The experimental setting is as follows.
First, we sample a random variable $x_1$ from a probability distribution $P(v)$, where $v\in\mathbb{R}$ is the variance.
We take three types of distributions as $P(v)$;
\[P(v):=\begin{cases}
    \frac{1}{\sqrt{2\pi v}}\exp\left(-\frac{x^2}{2v}\right) & (\mathrm{Gaussian~distribution}) \\
    \frac{v^x\exp(-v)}{x!} & (\mathrm{Poisson~distribution}) \\
    \frac{1}{\sqrt{2v}}\exp\left(-\frac{\sqrt{2}|x|}{\sqrt{v}}\right) & (\mathrm{Laplace~distibution}).
\end{cases}\]
Now, let us introduce a \textit{model function} $\varphi$, which describes the relationship between each node of input and output in a causal model.
For a model function $\varphi$, let $x_2=\varphi(x_1;e_2)$.
Here, we sample random variables $x_1$, and exogenous variables $e$ from $P(v)$ independently. %
As model functions, we set the following functions;
\[\varphi(x_1;e):=\begin{cases}
    cx_1+e & (\mathrm{Linear}) \\
    cx_1^2+cx_1+e & (\mathrm{Nonlinear\ polynomial}) \\
    \mathrm{sin}(cx_1+e) & (\mathrm{Nonlinear\ periodic}), 
\end{cases}\]
where the coefficient $c \in \{1, 10, 100\}$.
The coefficient directly affects the relative magnitude of values for linear and nonlinear polynomial models and the period for a nonlinear periodic model, but we observed similar behavior in all of them, hence we show results of $c=100$ in the main text, and the rest are found in Appendix.
Hereafter, we simply denote $\varphi(x_1)$ instead of $\varphi(x_1;e)$.
In addition to $x_1$ and $e$, we sample an independent random variable $x_3$ from $P(v)$.
The main purpose of this experiment is to determine the condition that the estimation of the mutual information with the quantum kernel method can discriminate that $x_3$ is independent of the other variables $x_1$ and $x_2$.

For a mutual information criterion $I(x,y)$ (see Sec.~\ref{section:intro}), we define an independence criterion
\[S(x_i):=\sum_{j=1}^3(1-\delta_{i,j})I(x_i,x_j).\]
This quantity is a measure of how independent a variable $x_i$ is compared to other variables.
This is also used to determine the order of causality in~\cite{shimizu2011directlingam}, which is the crucial point in LiNGAM.
In the setting of this experiment, $x_3$ is independent of $x_1$ and $x_2$, thus theoretically it should hold
\[S(x_3)=0.\]
In the following experiment, we will see whether the independence test succeeds or not by checking if the following relationship holds or not;  
\[S(x_3)<S_{\mathrm{min}}:=\mathrm{min}\left\{S(x_1), S(x_2)\right\}.\]

To estimate the quantity $S(x_i)$, we use the classical and the quantum kernels.
For the classical kernel method, we use the Gaussian kernel
\[k(x,y):=\exp\left(-\frac{||x-y||^2}{2}\right),\]
which corresponds to $\sigma = 1$ in \eqref{eq:rbf_kernel}.
The choice of the width $\sigma$ is due to the discussion in \cite{bach2002kernel}. They concluded that when the number of samples $N$ is less than $1000$, one might better use $\sigma=1$.
For the quantum kernel method, we use the following quantum circuit, similarly as~\cite{kawaguchi2023application}.
The data $x$ is encoded as 
\[U_{\mathrm{IQP}}(x)=H^{\otimes n}V_D(x)H^{\otimes n} \dots V_1(x)H^{\otimes n},\]
where $H$ is the Hadamard gate, and $V_i(x)(i=1, \cdots, D)$ consists of a polynomial number of diagonal matrices.
More concretely, $V_i(x)$ is constructed in two layers; the $U1$ gates are adapted to each all qubits, and then controlled-$U1$ gates are linearly adapted to each neighboring, where $U_1$ and controlled-$U_1$ gate are expressed in matrix form as follows, respectively;
\begin{equation*}
    U1(\lambda) = \begin{pmatrix} 1 & 0 \\ 0 & e^{i\lambda} \end{pmatrix},
    ~~~~\mathrm{controlled-}U1(\lambda) = \begin{pmatrix} 1 & 0 & 0 & 0 \\ 
                                                      0 & 1 & 0 & 0 \\
                                                      0 & 0 & 1 & 0 \\
                                                      0 & 0 & 0 & e^{i\lambda} \end{pmatrix}.
\end{equation*}

In addition to the variation of kernels, we check the influence of the activation function.
The activation function affects the sensitivity of the change in the input, and affects the task performance.
We introduced the following one for embedding $x$ into the kernel, which was introduced in \cite{kawaguchi2023application};
\[\mathrm{tanh\mathchar`-shrink}(x)=x-\mathrm{tanh}(x).\]

For more details, see Sec. \ref{section:kernel}.

\subsection{Experimental results}
\subsubsection{Comparison between quantum and classical kernels}
First, we take the normal distribution with $v=1$ as $P(1)$.
Table \ref{table:result_kernels} shows the results of the independence test performed with classical kernel and quantum kernels.
Here, we also list the results of the case with and without the activation function.
The values represent the correct ratio per 100 questions, which are generated with 100 different random seeds, for different sample sizes $N$.
It can be seen that the accuracy percentage of the case with the activation function is higher than that of the case without it for all sample sizes, regardless of the model function.
This result implies that the data embedding with activation functions is superior for the quantum kernel methods to estimate MI.
We designed the activation function while considering the continuity and the symmetricity of embedding space. 
In the case of the quantum kernel, data are embedded as the phase of a quantum state, which is uniformly continuous, hence we chose a symmetric and continuous type activation function to evenly utilize the Hilbert space with positive and negative input values.
The experimental result seems to support our insight, but the activation function is a result of engineering and future theoretical analysis is desirable.

Another notable result is the performance of ``Nonlinear periodic" case.
Classical kernel does not work well at all, but quantum kernel shows some capability of answering correct dependency.
This simply seems due to the compatibility of the quantum kernel method with periodic data. 
This is a well-known fact about quantum machine learning~\cite{PhysRevA.103.032430}.
In the real world, there exist various phenomena that output periodical data or following periodical rules, hence quantum kernel may be suitable for analyzing such datasets.

\begin{table}[htb]
\caption{The correct ratio of the independence test for a normal distribution. ``with (without)" denotes ``with (without) activation function". (unit: \%, $v=1$, $c=100$)}
\label{table:accuracy_gaussian}
  \begin{minipage}[t]{.49\textwidth}
    \begin{center}
    \subcaption{Estimation of MI.}
    \scalebox{0.9}{
      \begin{tabular}{|c|c||c|c|c|c|c|c|}
    \hline
\multicolumn{1}{|c|}{\multirow{2}{*}{Model}} & \multicolumn{1}{c||}{\multirow{2}{*}{$N$}} & \multicolumn{1}{c|}{\multirow{2}{*}{Classical}} & \multicolumn{2}{c|}{Quantum}                             \\ \cline{4-5} 
\multicolumn{1}{|c|}{} & \multicolumn{1}{c||}{} & \multicolumn{1}{c|}{}                           & \multicolumn{1}{c|}{~~with~~} & \multicolumn{1}{c|}{without} \\
      \hline\hline
      \multirow{3}{*}{Linear} & 10& 48 &41 &32\\ 
     & 30  & 44 &58  &28\\ 
     & 50  & 66& 72&35 \\ \hline
          \multirow{3}{5em}{Nonlinear polynomial} & 10&  57& 46 &32\\ 
     & 30  &68  &92 &33\\
     & 50  & 90&100 &50 \\ \hline
          \multirow{3}{5em}{Nonlinear periodic} & 10& 0 & 38 &27\\
     & 30  &0 &37 &37\\
     & 50  & 0 & 57&35\\ \hline
      \end{tabular}}
    \end{center}
  \end{minipage}
  \begin{minipage}[t]{.49\textwidth}
    \begin{center}
    \subcaption{Estimation of SMI.}
     \scalebox{0.9}{     \begin{tabular}{|c|c||c|c|c|c|c|c|}
    \hline
\multicolumn{1}{|c|}{\multirow{2}{*}{Model}} & \multicolumn{1}{c||}{\multirow{2}{*}{$N$}} & \multicolumn{1}{c|}{\multirow{2}{*}{Classical}} & \multicolumn{2}{c|}{Quantum}                             \\ \cline{4-5} 
\multicolumn{1}{|c|}{} & \multicolumn{1}{c||}{} & \multicolumn{1}{c|}{}                           & \multicolumn{1}{c|}{~with~~} & \multicolumn{1}{c|}{without} \\
      \hline\hline
      \multirow{3}{*}{Linear} & 10&38  &26&36\\
     & 30  & 53 &33 &32\\
     & 50  &46 &35&27 \\ \hline
          \multirow{3}{5em}{Nonlinear polynomial} & 10&52 &38 &32\\
     & 30  & 54 &35&39\\
     & 50  & 50 &38&29 \\ \hline
          \multirow{3}{5em}{Nonlinear periodic} & 10 & 0&33&29\\
     & 30  & 0  &28&39\\
     & 50  &0 &  38&29\\ \hline
      \end{tabular}}
    \end{center}
  \end{minipage}
  \label{table:result_kernels}
\end{table}

\subsubsection{Comparison of different variances $v$}
\begin{figure}[htbp]
  \begin{minipage}[b]{0.32\hsize}
    \centering
    \includegraphics[keepaspectratio, scale=0.36]{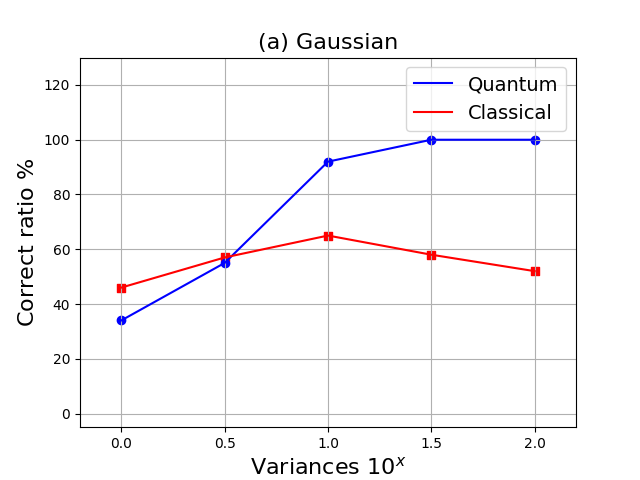}
  \end{minipage}
  \begin{minipage}[b]{0.32\hsize}
    \centering
    \includegraphics[keepaspectratio, scale=0.36]{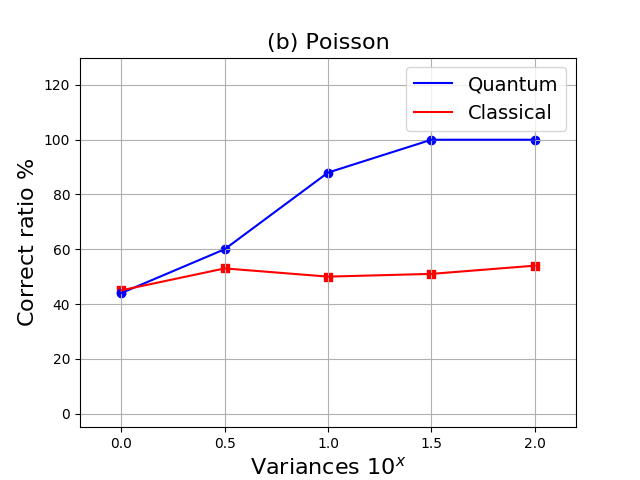}
  \end{minipage}
  \begin{minipage}[b]{0.32\hsize}
    \centering
    \includegraphics[keepaspectratio, scale=0.36]{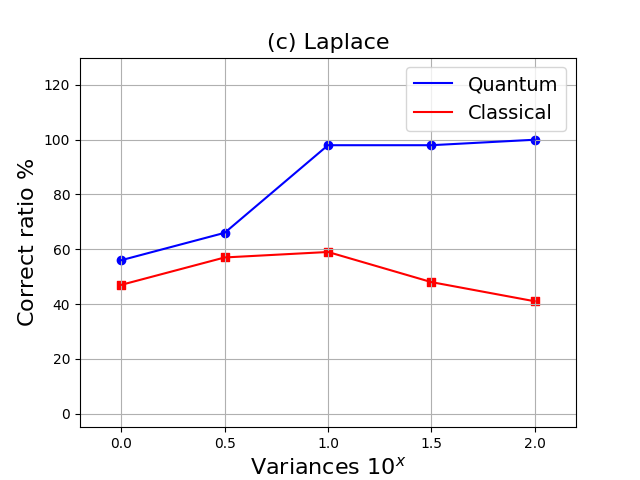}
  \end{minipage}
  \begin{minipage}[b]{0.32\hsize}
    \centering
    \includegraphics[keepaspectratio, scale=0.36]{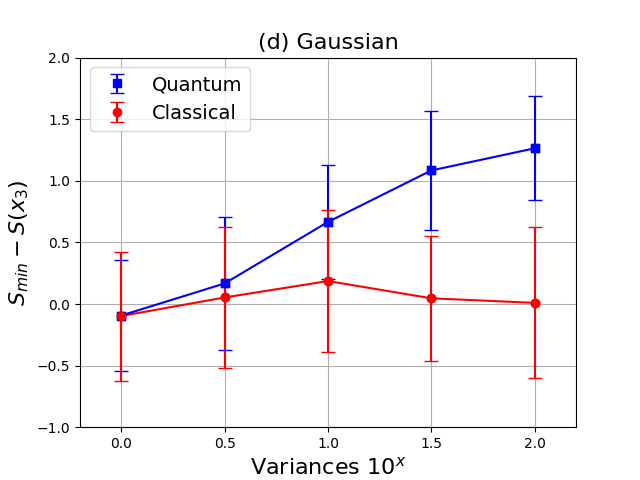}
  \end{minipage}
  \begin{minipage}[b]{0.32\hsize}
    \centering
    \includegraphics[keepaspectratio, scale=0.36]{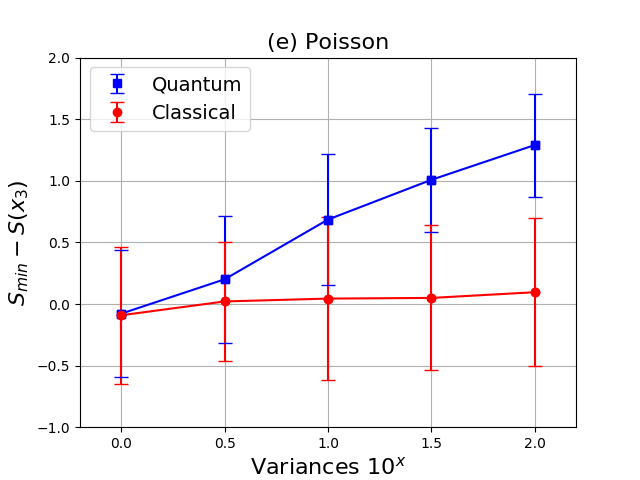}
  \end{minipage}
  \begin{minipage}[b]{0.32\hsize}
    \centering
    \includegraphics[keepaspectratio, scale=0.36]{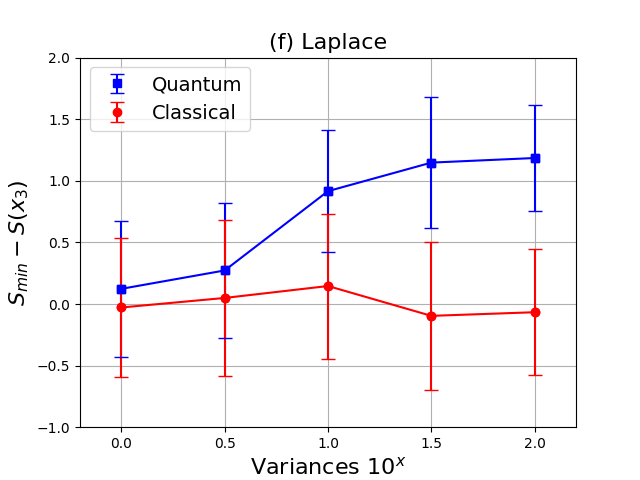}
  \end{minipage}
    \caption{The results of the independence test with different variances. (Model: Linear, $c=100, N=10$)
(a-c) represent the correct ratio, and (d-f) represents the slack values. The probable distribution is described in each figure.}
    \label{figure:slack_diff_var}
\end{figure}

Next, we compare the performances between classical and quantum kernels for the data sets that have different variances.
FIG. \ref{figure:slack_diff_var} (a-c) shows the correct ratio of the independence test 
based on MI with different variances for each distribution performed with classical and quantum kernels, and the causal model is the linear relational equations.
It is clearly observed that the quantum kernel shows a higher correct ratio than the classical kernel as increasing the variance for any distribution.
The accuracy of MI estimated with the classical kernel is lower than that estimated with the quantum kernel except in the range where the variance is quite small.
In addition, the performance of the classical kernel decreases as the variance of the given data is far from $v=1$, which corresponds with the variance of the 
 Gaussian kernel.
On the other hand, the quantum kernel keeps improving the estimation accuracy as the variance of given data becomes larger.
We can interpret these results based on the shape of kernels.
The absolute value of the kernel heavily affects the value of the Gram matrix.
More concretely, if the value is quite small, one can not detect the changes in the data.
The Gaussian kernel has a shape of $\exp(-x^2)$, which decreases rapidly to zero as $x$ becomes large.
Therefore, it could have difficulty to discriminate data with large variances.
In fact, in the case that a dataset has a sufficiently large variance, the Gram matrices corresponding to all variables approach the identity matrix, indicating that the mutual information cannot be distinguished.
On the other hand, the quantum kernel, which is constructed of the output from quantum circuits, could have a wide tail compared to the classical kernel due to the property of anti-concentration, as represented by the theory of random walks~\cite{dalzell2022random}.
We need theoretical work more in detail, but it is qualitatively reasonable that the quantum kernel would be a superior estimator for non-Gaussian probability distributions.

FIG. \ref{figure:slack_diff_var} (d-f) shows that the slack values, calculated by $S_{\rm min}-S(x_3)$. %
The dots show the mean value, and the width of the error bars corresponds to a standard deviation obtained by 100 trials.
Note that the number of samples in this experiment is 10, and the model function is linear.
The larger slack values correspond to better performance, and the negative slack value means that the estimation is a failure, hence these results coincide with FIG.~\ref{figure:slack_diff_var} (a-c).
This result clearly shows the fact that the quantum method offers a better answer than the classical one, since the error bars for estimation errors using the classical and quantum kernel methods no longer overlap as the variance is increased.

\subsubsection{Comparison of different sample size $N$}
\begin{figure}[htbp]
  \begin{minipage}[b]{0.48\hsize}
    \centering
    \includegraphics[keepaspectratio, scale=0.48]{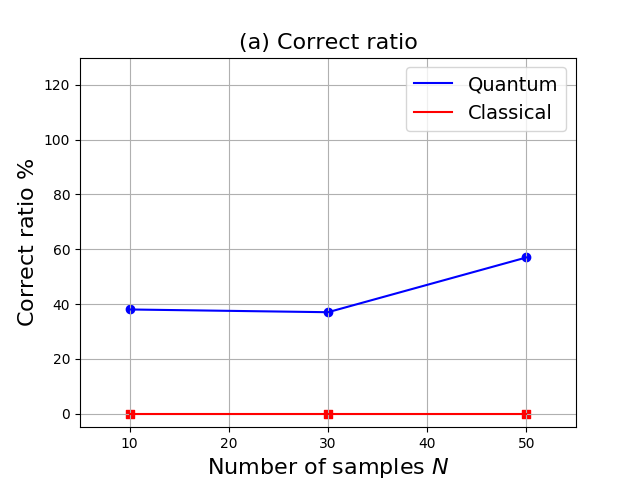}
  \end{minipage}
  \begin{minipage}[b]{0.48\hsize}
    \centering
    \includegraphics[keepaspectratio, scale=0.48]{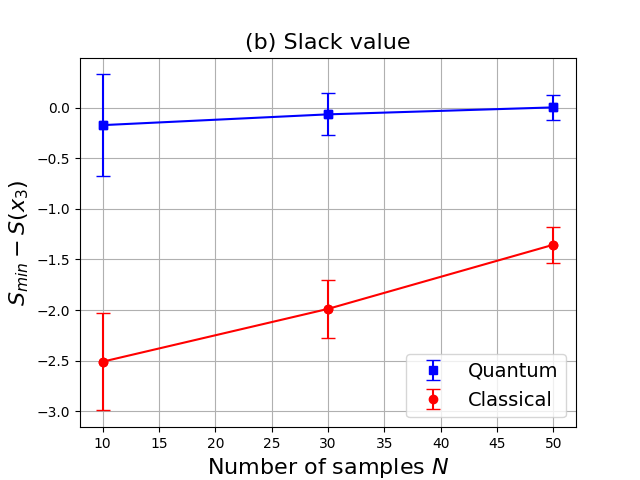}
  \end{minipage}
  \caption{The results of the independence test with a different number of samples. ($c=100, v=1$, Model: Nonlinear periodic, Gaussian)}
    \label{figure:sin_diff_sample}
\end{figure}

Finally, we check the performance for a different number of samples.
We focus on the data sampled from the periodic causal model, in which the quantum kernel shows a clear advantage compared to the classical kernel in TABLE~\ref{table:accuracy_gaussian}.
The experimental results are displayed in FIG.~\ref{figure:sin_diff_sample}.
It is observed from (b) that the slack values for both kernels improve as the sample number becomes large.
However, the slack value of the classical kernel method is negative even with 50 samples, meaning failure in estimating the correct independence.
In contrast to that, although the slack value of the quantum kernel is negative when the sample number is small, it turns over to positive at the sample number of 50, and the correct ratio of the quantum kernel shows over 50$\%$ at $N=50$ in  FIG.~\ref{figure:sin_diff_sample}~(a).%
To interpret this result, we need a more detailed analysis, but the advantage of the quantum kernel could be due to the fact that the central limit theorem does not hold under the small sample number.
It is empirically known that the probability distribution generated by the small number of samples does not follow the central limit theorem and produces certain outliers when $N<30$.
This point has been classically investigated in statistics~\cite{israel1992determining, chang2006determination}, and a similar analysis of the quantum kernel method is desirable.
The classical kernel method (with Gaussian kernel) cannot handle such outliers appropriately, but the quantum kernel method could thanks to the non-classical nature (such as the anti-concentration).
This phenomenon, anti-concentration, has been established as a theorem concerning fundamental aspects of quantum circuit design \cite{dalzell2022random, hangleiter2018anticoncentration}, in addition to quantum random walks. 
It is interesting to apply these theorems to the circuits and the activation functions.

\section{Discussion}
\label{section:analysis}
This section provides a theoretical discussion of the experiments conducted in Sec.\ref{section:experiments}. We consider that our experimental results on MI can be explained by the scope of the central limit theorem; see Subsec. \ref{subsection:connection to the central limit theorem}.
On the other hand, the try to estimate SMI using quantum kernel methods seems to fail.
We will discuss this phenomenon in Subsec. \ref{subsection:the structure of RKHS}.

\subsection{Connection to the central limit theorem}
\label{subsection:connection to the central limit theorem}
We discuss the quantum advantage observed under small sample sizes, large variances and nonlinear model functions.
One of the possible factors that cause the advantages could be the effect of the central limit theorem and anti-concentration, which is a unique property of quantum models.
The main concept of the central limit theorem is that if the sample size is large enough, the probability distribution of those averages will closely approximate a normal distribution \cite{feller1991introduction}.
In particular, as is well known, if the hem of the distribution followed by the probability distribution is sufficiently large, it will converge to a stable distribution with characteristic exponent $0<\alpha<2$ rather than a normal distribution. In this case, the variance becomes infinite and has a fat tail, which is similar to our case \cite{voit2002statistical}.
Although it holds for any sample distribution, the convergence rate of course depends on the original distribution.
That is, the more complex distribution, meaning that the deviation from the normal distribution is large, needs more samples.
In our case, the probability distributions from which variables sampled include non-Gaussian and the model functions have different non-linearity, therefore the observed behavior largely reflects on such differences.
More concretely, the performances of the quantum method are comparable with the classical one or worse than that in the case of linear models even under a small sample size.
On the other hand, the quantum method shows better results with higher-order polynomials and periodic equations.
These results can be interpreted that several dozen samples are enough to qualitatively approximate normal distribution for simple linear models, and the classical method could straightforwardly work, but for more complex model functions such as higher order polynomials and periodic functions, such sample size is not enough to appropriately work for the classical method.

As observed in FIG.~\ref{figure:slack_diff_var} (b), in the case of periodic functions,  even the case of $N = 50$ can also be considered a small sample.
In this light, instead of stating the advantage in terms of periodicity, it may be better to make a quantitative evaluation of the central limit theorem in terms of the complexity of the model function (e.g. degree of the polynomial equation) and combine it with anti-concentration.
Of course, by the central limit theorem, the compensated approximation accuracy depends on the complexity of the data. Therefore, the experiments in this paper reflect them in terms of the kernel method.

\subsection{The structure of RKHS} %
\begin{figure}[htbp]
  \begin{minipage}[b]{0.48\hsize}
    \centering
    \includegraphics[keepaspectratio, scale=0.48]{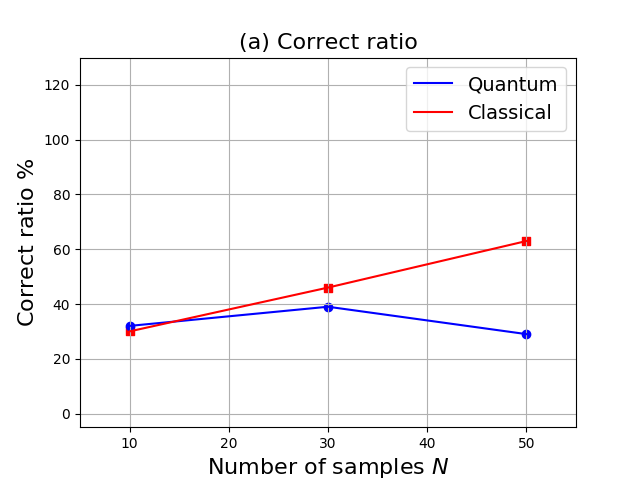}
  \end{minipage}
  \begin{minipage}[b]{0.48\hsize}
    \centering
    \includegraphics[keepaspectratio, scale=0.48]{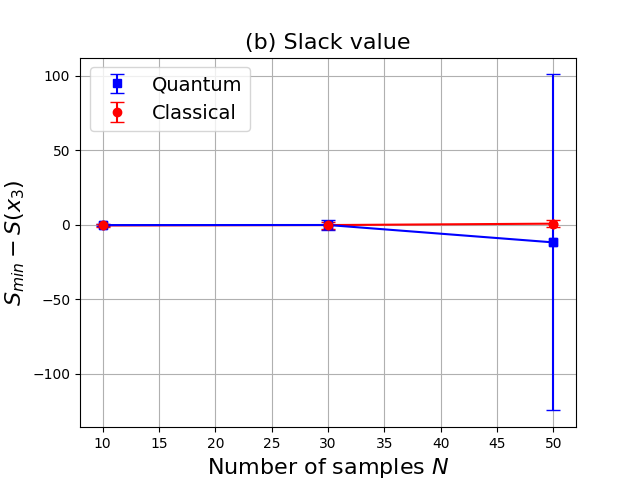}
  \end{minipage}
  \caption{The results of the independence test with a different number of samples. ($c=100, v=1$, Model: Linear, Poisson)}
    \label{figure:inocco_diff_sample}
\end{figure}

\label{subsection:the structure of RKHS}
We introduce the mathematics of the structure of the RKHS. %
As mentioned in Sec.~\ref{section:intro}, SMI has several merits compared to MI, and 
we tried to estimate SMI with quantum kernels.
However, we have experimentally observed an unexpected behavior such that the slack value shows large blurring as the number of samples is increased, i.e. FIG. \ref{figure:inocco_diff_sample}~(b).
To interpret this behavior, let us recall the assumption required to estimate SMI by kernel method.
As mentioned in Sec.~\ref{subsection:smi}, the quantity \eqref{eq:smi_estimated} is a good of SMI under the assumption that the kernel satisfies the property of characteristic.
Under that condition, it is known that the estimator coincides with SMI in the limit of a large number of samples, independently of the specific form of a kernel ~\cite{fukumizu2007kernel}.
With considering the experimental result and above mentioned assumption, we suspect that the quantum kernel, or the kernel composed of IQP circuit, is not characteristic.
We have not reached a conclusion yet, but the summary of current knowledge regarding that is as follows.
Steinwart introduced the notion of \textit{universality} of a kernel and showed that the universal kernel is characteristic~\cite{steinwart2001influence}.
As an application, he proved that the Gaussian kernel is universal and hence characteristic.
As he computed, whether the kernel is universal or not can be determined from each coefficient in the Taylor expansion of the kernel.
However, it is difficult to Taylor expand the quantum kernel and see the coefficients in our case, particularly because of the activation function.
As a relational work, Schuld-Killoran~\cite{schuld2019quantum} studied the quantum kernel in explicit form, hence it could be possible to perform such a calculation.

\section{Application of the estimation of mutual information and the independence test}
\label{section:application}
In this section, we review the application of the estimation of mutual information and the independence test.
They have wide applications such as feature selection \cite{fang2015feature}, clustering \cite{slonim2005information}, gene networking \cite{jiang2022gene} and anomaly detection \cite{kopylova2008mutual}, and here we focus on the causal discovery, which is one of the most important application.
Several methods are known for the causal discovery, depending on whether the relationship between nodes is linear or nonlinear, and whether the assumptions of the model to be estimated are (semi-)parametric or non-parametric.
The oldest and the most widely used non-parametric method is the PC algorithm~\cite{richardson2013polynomial, spirtes1991algorithm, spirtes2013causal}.
This algorithm is proposed in \cite{spirtes2000causation}, and it determines the directions between each pair of variables recursively. 
On the other hand, LiNGAM, which has been mentioned in this paper, is a semi-parametric method~\cite{shimizu2011directlingam}. 
There are also methods that extend this to nonlinear or mention the existence of hidden variables~\cite{hoyer2008estimation, lacerda2012discovering, shimizu2014lingam, shimizu2016non, zhang2016nonlinear}.
The method qLiNGAM is also this kind of extension~\cite{kawaguchi2023application}.
All aforementioned methods use an independence test to determine the order of causality between nodes.

If we consider the practical application, we also need to consider the calculation cost.
Particularly, the computational complexity of the estimation method with (quantum) kernel becomes highly demanding as increasing the sample number $N$, at least $O(N^2)$.
To overcome that problem, we need to introduce an approximation method for calculating the kernel in practice.
In the classical case, various scalable kernel methods have been proposed~\cite{Rahimi2007-om,williams2000using,smola2000sparse,hsieh2014divide,zhang2013divide,liu2020learning}. 
Among them, the most popular framework is the method using random Fourier features~\cite{Rahimi2007-om};
The idea is to approximate the target kernel by the inner products of probabilistically generated feature vectors, where these two quantities are guaranteed to be close
with each other.
Also, in the quantum case, a scalable framework is proposed to estimate the target kernel efficiently~\cite{nakaji2022deterministic}.

\section{Conclusion and future work}
\label{section:future_work}
This paper introduces a kernel-based method for estimating mutual information and compares the performance of the classical and quantum kernel methods. Our experimental results show that the quantum kernel method better determines the independence between data than the classical one in situations where the variance is large or only small samples are available.
In these cases, the distribution of the data cannot be approximated by a normal distribution. Similar results are also found for data with periodic functions as relations.

Below, we list the remaining open problems that we have not fully addressed in this paper.
The first one is the mathematics of quantum kernel methods.
The problems are as follows.
\begin{enumerate}
    \item It is known that, unlike the classical situations, the resulting probability distribution has a large base in a random walk using quantum circuits~\cite{dalzell2022random}. In this study, we saw that the performance of the quantum kernel method improves with respect to probability distributions with large variance. Can this be explained in terms of anti-concentration?
    \item For the estimation of SMI, we used a quantum kernel method consisting of the IQP circuits. However, as we increased the number of samples, we observed that the numerical variance of the experimental results increased. 
    Is the IQP circuit characteristic or more strongly universal?
    How can we describe the class of quantum kernels that are characteristic?
    \item It is known that the performance of kernel methods is measured by the Rademacher complexity~\cite{cortes2013learning}. Is it possible to characterize quantum kernels by such a measure?
    \item To develop a learning theory for kernel methods, it is important to discuss convergence rates, combined with the uniformity in function approximation for output functions.
    The most famous of these is the Barron class, describing the approximation and estimation theory initiated by Barron~\cite{barron1993universal, barron1994approximation}.
    Can we explicitly describe the function classes realized by quantum kernel methods in this perspective?
    This is expected to be one of the explanations for the advantage over small samples.
\end{enumerate}

The third one is the quantum causal discovery.
In the context of the causal discovery, we impose a Markov property on the graph.
Quantum mechanics involves difficult issues regarding state transitions, and there exists the notion of a quantum Markov process~\cite{barry2014quantum, cholewa2017quantum, monras2010hidden}.
Causal discovery is one useful application that has been successful using graphical models, but it can be extended quantum mechanically.
In particular, it has been shown that it is possible to fully recover a model from quantum data in some cases~\cite{brukner2014quantum, ried2015quantum}, and quantum causal discovery has also been actively studied~\cite{costa2016quantum, leifer2008quantum}.
It would be an interesting problem to compare the performance of those and the method of this paper~\cite{pearl2009causality}.

\begin{acknowledgments}
This work was supported by MEXT Quantum Leap Flagship Program Grant Number JPMXS0118067285 and JPMXS0120319794.
\end{acknowledgments}

\bibliographystyle{abb_srt}
\bibliography{main}

\newpage
\appendix
\section{Complementary experiments}

\subsection{Results with Gaussian noise model}
\begin{figure}[htbp]
      \begin{minipage}[b]{0.32\hsize}
    \centering
    \includegraphics[keepaspectratio, scale=0.36]{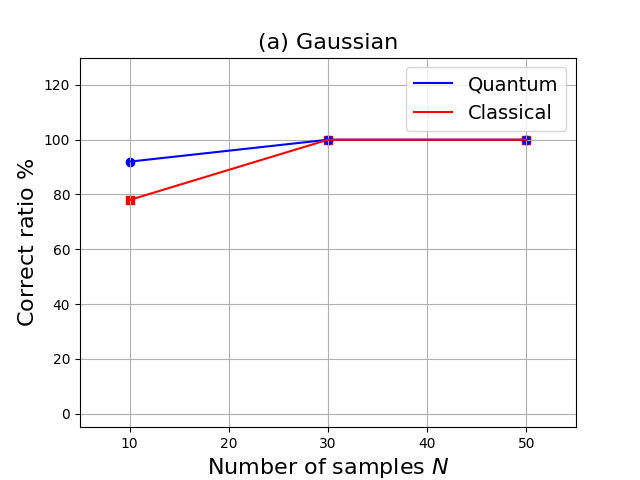}
  \end{minipage}
  \begin{minipage}[b]{0.32\hsize}
    \centering
    \includegraphics[keepaspectratio, scale=0.36]{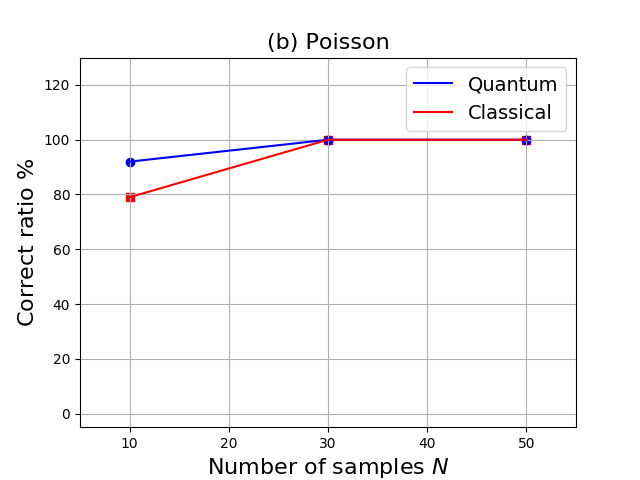}
  \end{minipage}
  \begin{minipage}[b]{0.32\hsize}
    \centering
    \includegraphics[keepaspectratio, scale=0.36]{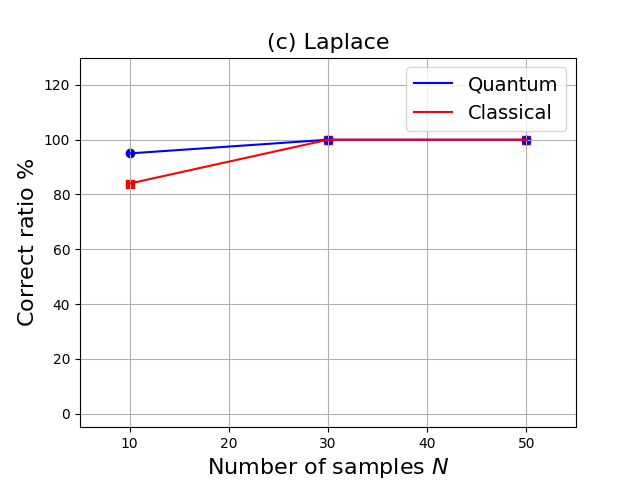}
  \end{minipage}
  \begin{minipage}[b]{0.32\hsize}
    \centering
    \includegraphics[keepaspectratio, scale=0.36]{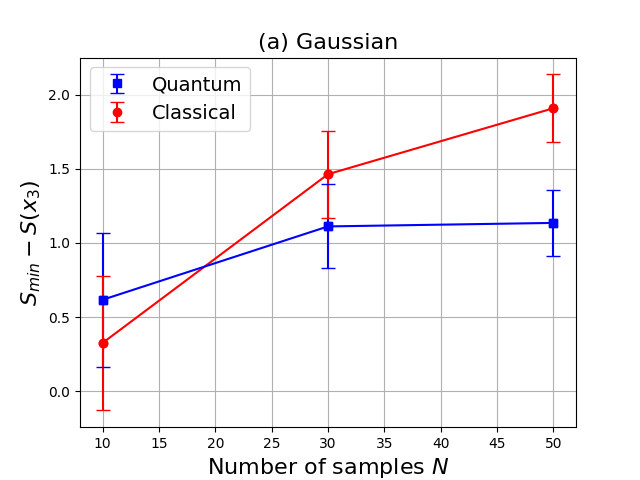}
  \end{minipage}
  \begin{minipage}[b]{0.32\hsize}
    \centering
    \includegraphics[keepaspectratio, scale=0.36]{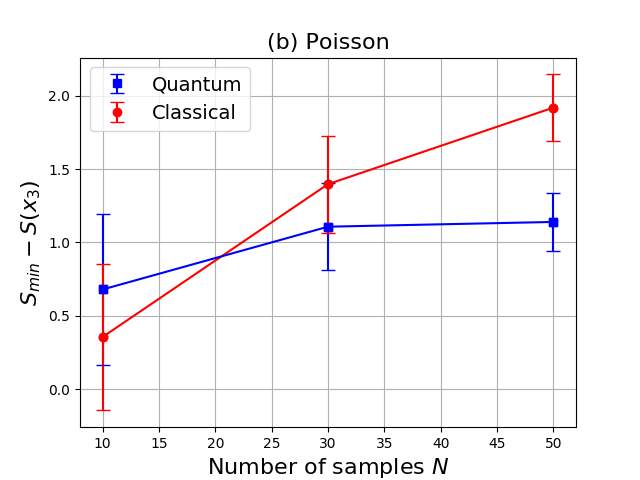}
  \end{minipage}
  \begin{minipage}[b]{0.32\hsize}
    \centering
    \includegraphics[keepaspectratio, scale=0.36]{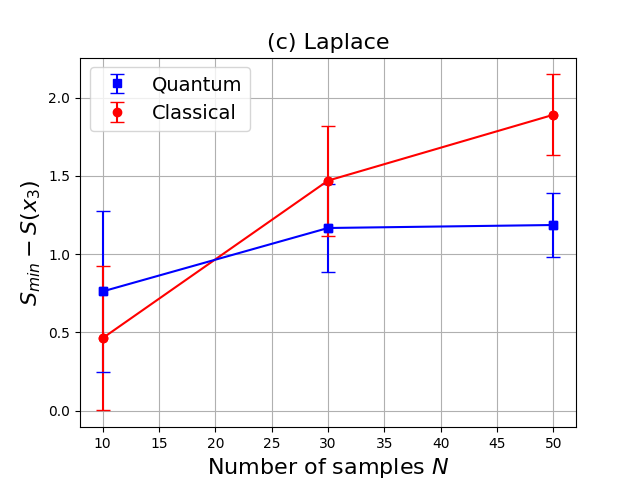}
  \end{minipage}
  \caption{The results of the independence test with different sample sizes. (Noise: Gaussian, $c=100, v=1$)
(a-c) represent the correct ratio, and (d-f) represents the slack values. The probable distribution is described in each figure. (Model: Linear)}
  \label{appfigure:accuracy_diff_var}
\end{figure}

In this section, we describe the experimental results with different noise models to the main text's.
We set the probability distribution of the noise $e$ appearing in the model function $\varphi(x_1;e)$ as a normal distribution with mean 0 and variance 1, which are $P(v)$ in the main text.
We take the exogenous variables $x_1$ from the probability distribution $P(1)$ in this section.
The experimental results with the model function ``Linear" are shown in Fig. \ref{appfigure:accuracy_diff_var}.
In the result of the correct ratio, both the quantum and the classical method show $100 \%$ at $N=30, 50$ with all model functions, and the slack values also support such results, meaning the values are sufficiently large at those points.
However, as mentioned in the main text, the quantum kernel method performs better than the classical kernel method when the number of samples is 10.
Therefore, although we cannot address all possible settings, we consider that the advantage that the quantum kernel method represents under a small sample size is not a coincidence but an intrinsic nature of the quantum kernel.

\subsection{Results with other parameter settings}
In this section, we describe the complementary experimental results.
We conduct the same experiments represented in the main text with various different parameter settings to see the overview.

\subsubsection{Sample size dependence}
\begin{figure}[htbp]
  \begin{minipage}[b]{0.32\hsize}
    \centering
    \includegraphics[keepaspectratio, scale=0.36]{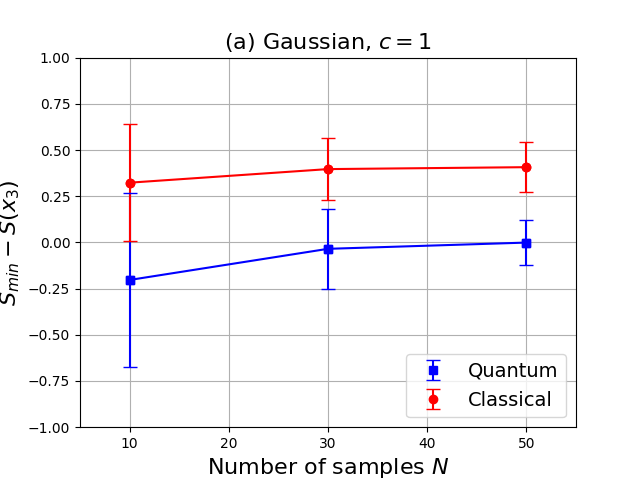}
  \end{minipage}
  \begin{minipage}[b]{0.32\hsize}
    \centering
    \includegraphics[keepaspectratio, scale=0.36]{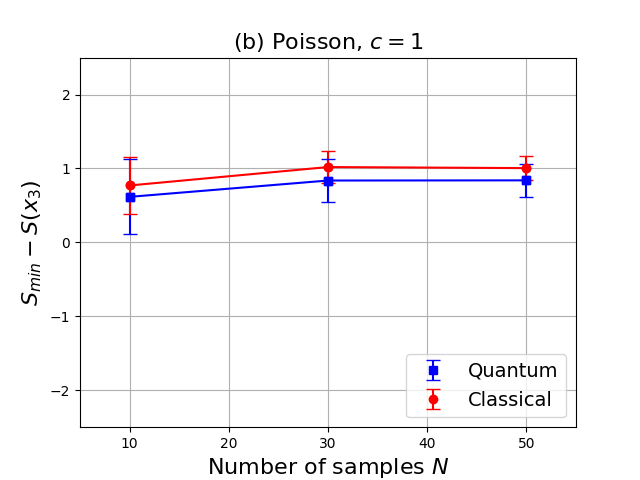}
  \end{minipage}
  \begin{minipage}[b]{0.32\hsize}
    \centering
    \includegraphics[keepaspectratio, scale=0.36]{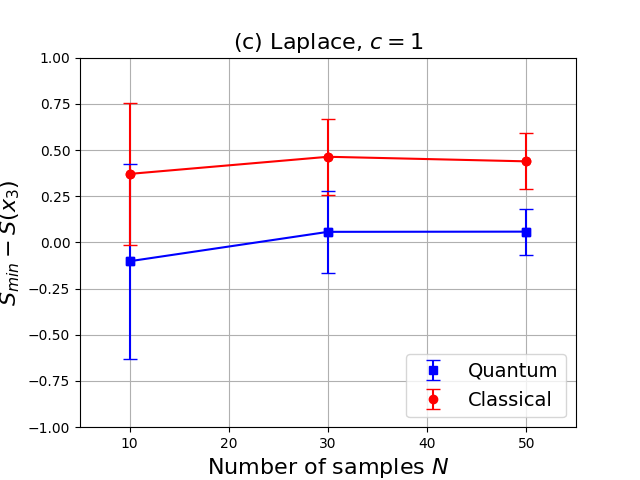}
  \end{minipage}
  \begin{minipage}[b]{0.32\hsize}
    \centering
    \includegraphics[keepaspectratio, scale=0.36]{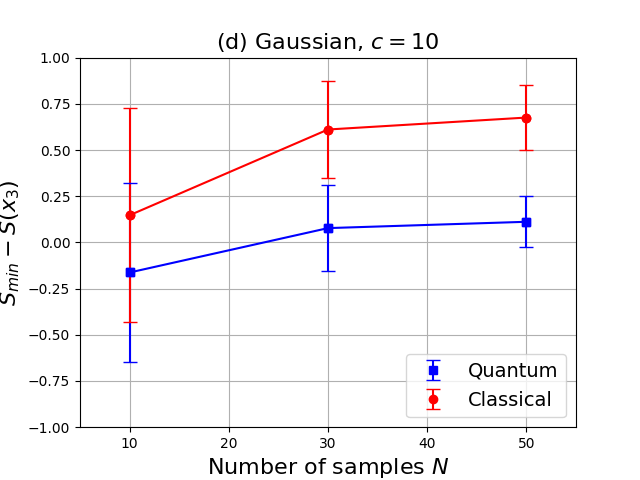}
  \end{minipage}
  \begin{minipage}[b]{0.32\hsize}
    \centering
    \includegraphics[keepaspectratio, scale=0.36]{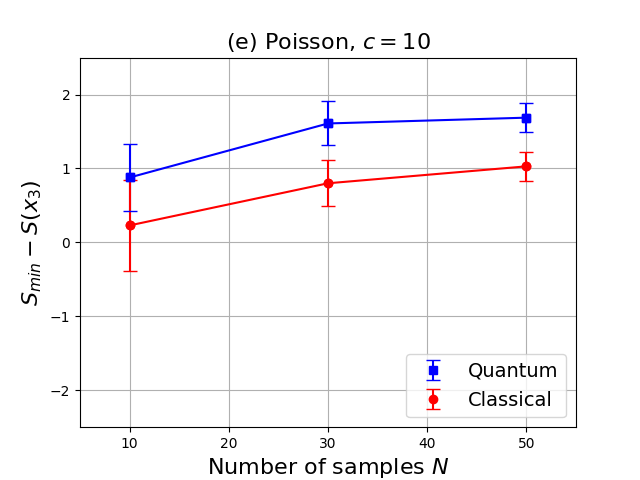}
  \end{minipage}
  \begin{minipage}[b]{0.32\hsize}
    \centering
    \includegraphics[keepaspectratio, scale=0.36]{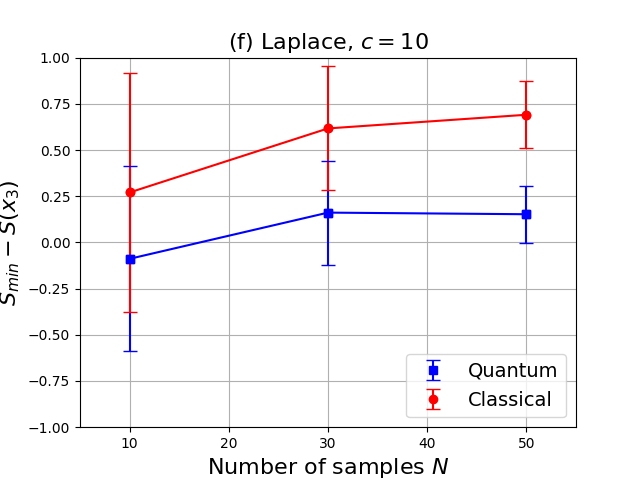}
  \end{minipage}
  \begin{minipage}[b]{0.32\hsize}
    \centering
    \includegraphics[keepaspectratio, scale=0.36]{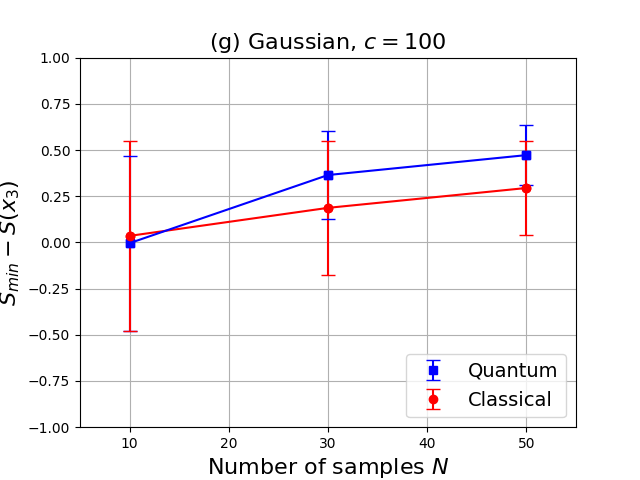}
  \end{minipage}
  \begin{minipage}[b]{0.32\hsize}
    \centering
    \includegraphics[keepaspectratio, scale=0.36]{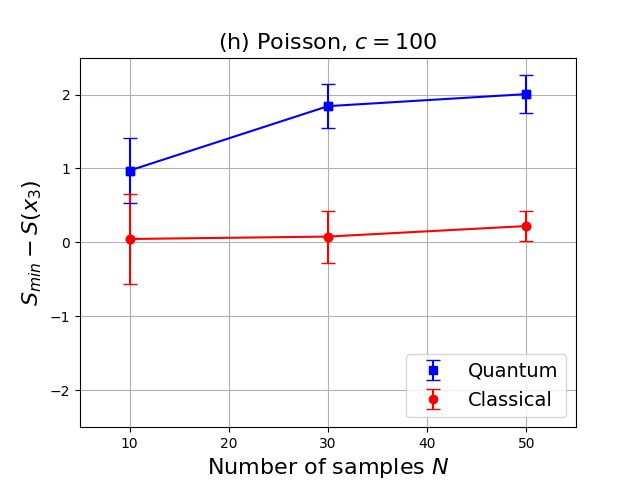}
  \end{minipage}  
  \begin{minipage}[b]{0.32\hsize}
    \centering
    \includegraphics[keepaspectratio, scale=0.36]{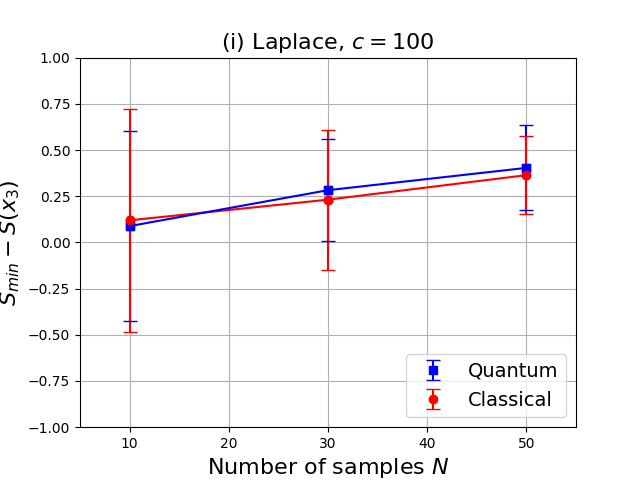}
  \end{minipage}
  \caption{The slack values of the independence test with different sample sizes. (Model: Nonlinear polynomial, $v=1$)}
  \label{figure:error_diff_samples_c_100}
\end{figure}

\begin{figure}[htbp]
  \begin{minipage}[b]{0.32\hsize}
    \centering
    \includegraphics[keepaspectratio, scale=0.36]{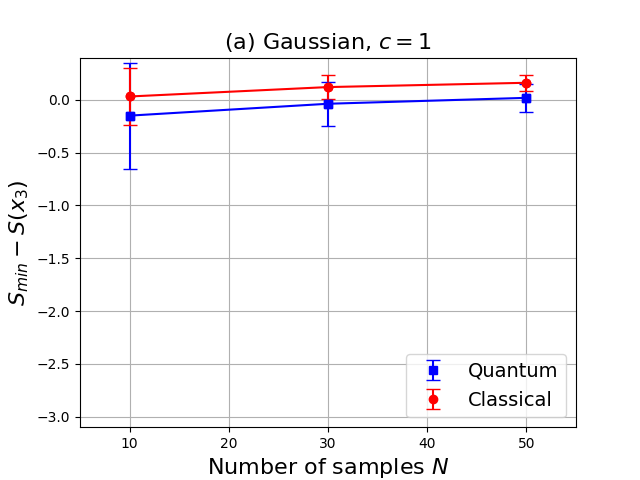}
  \end{minipage}
  \begin{minipage}[b]{0.32\hsize}
    \centering
    \includegraphics[keepaspectratio, scale=0.36]{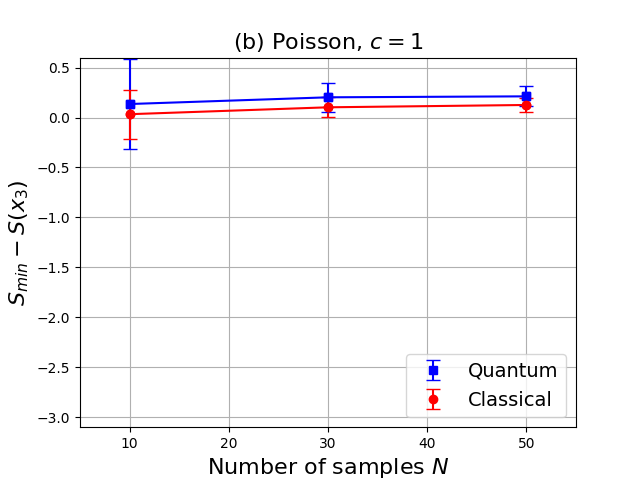}
  \end{minipage}
  \begin{minipage}[b]{0.32\hsize}
    \centering
    \includegraphics[keepaspectratio, scale=0.36]{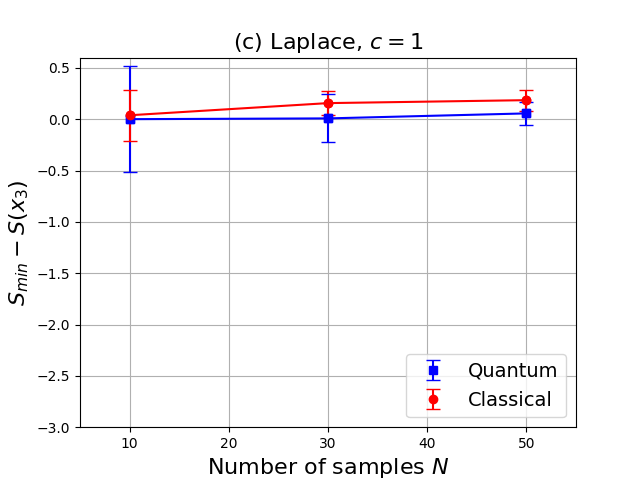}
  \end{minipage}
  \begin{minipage}[b]{0.32\hsize}
    \centering
    \includegraphics[keepaspectratio, scale=0.36]{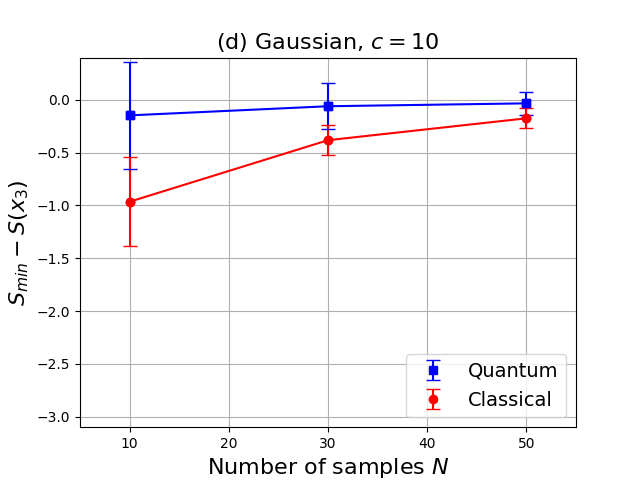}
  \end{minipage}
  \begin{minipage}[b]{0.32\hsize}
    \centering
    \includegraphics[keepaspectratio, scale=0.36]{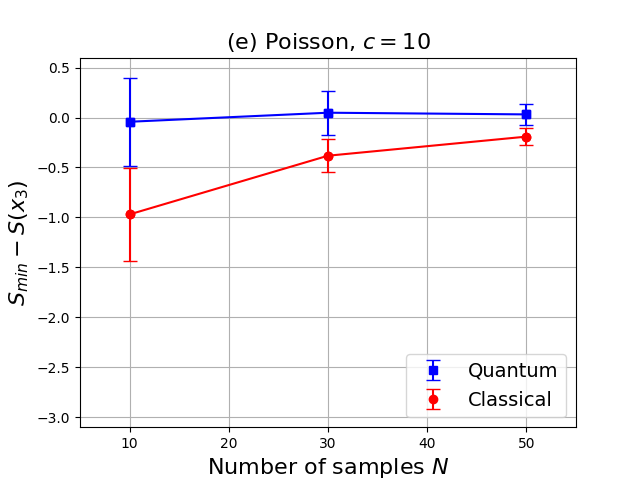}
  \end{minipage}
  \begin{minipage}[b]{0.32\hsize}
    \centering
    \includegraphics[keepaspectratio, scale=0.36]{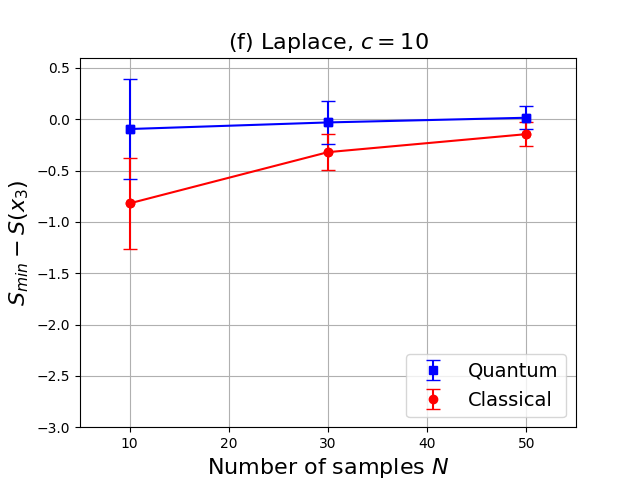}
  \end{minipage}
  \begin{minipage}[b]{0.32\hsize}
    \centering
    \includegraphics[keepaspectratio, scale=0.36]{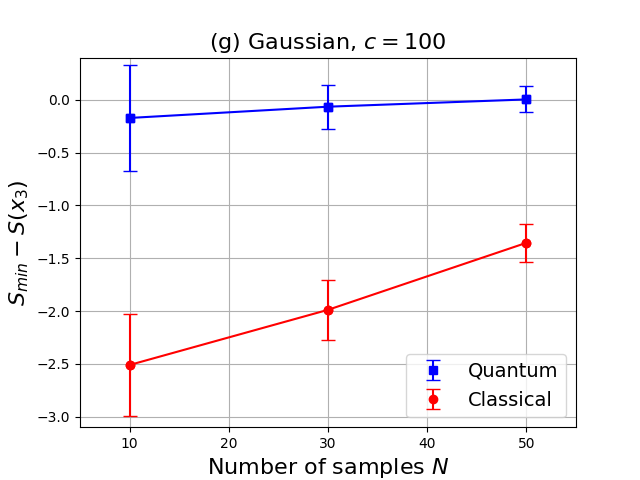}
  \end{minipage}
  \begin{minipage}[b]{0.32\hsize}
    \centering
    \includegraphics[keepaspectratio, scale=0.36]{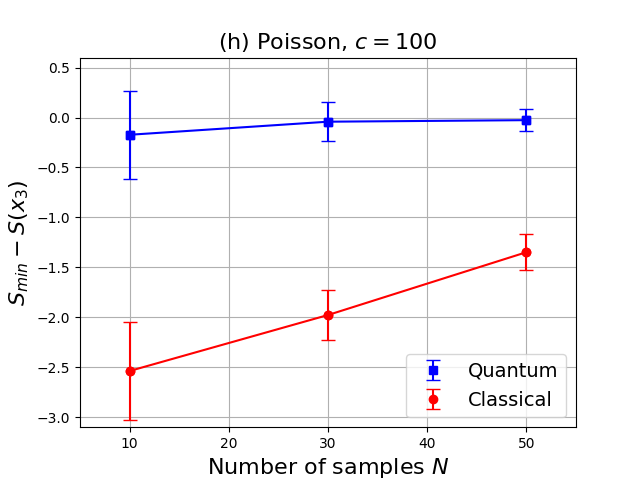}
  \end{minipage}  
  \begin{minipage}[b]{0.32\hsize}
    \centering
    \includegraphics[keepaspectratio, scale=0.36]{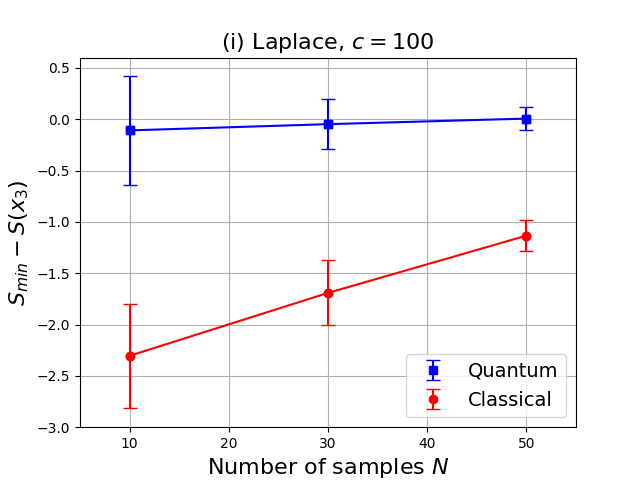}
  \end{minipage}
  \caption{The slack values of the independence test with different sample sizes. (Model: Nonlinear periodic, $v=1$)}
  \label{figure:error_diff_samples_c_1}
\end{figure}

FIG. \ref{figure:error_diff_samples_c_100} and \ref{figure:error_diff_samples_c_1} represent the sample size dependence of slack values with the model function ``Nonlinear polynomial" and ``Nonlinear periodic", respectively.
The coefficient in the model function $\varphi$ is not only 100, as in the main text, but also 1, 10.
It can be seen that for both non-linear polynomial and periodic functions, the estimation by the quantum kernel method has an advantage as the value of the coefficients is increased. 
This behavior would be due to the relative magnitude of values for nonlinear polynomial models and the period for a nonlinear periodic model.
For instance, in FIG. \ref{figure:error_diff_samples_c_1} (a-c; $c=1$), both the quantum and the classical method show similar performance, meaning no quantum advantage is observed, while the quantum one shows clearly better performance in (g-i; $c=100$).
Here, note that the coefficient in the nonlinear periodic function corresponds to the frequency of the trigonometric function, hence it has only small non-linearity if the coefficient is small.
Therefore, those results imply that the quantum kernel fits better for highly non-linear functions.
These results suggest when we should use the quantum kernel method rather than the classical one. 
That is, when only small samples are available and/or the model functions have complex nonlinearities. 
However, to interpret whole behaviors is not easy, hence we rest it as a future task.
Furthermore, the results suggest more important facts about the compatibility of probability distributions and kernel methods.
As can be seen by comparing the figures with the probability distribution $P(v)$, the quantum kernel method is suitable for the Poisson distribution, followed by the normal distribution and finally the Laplace distribution, in that order. 
Quantifying the complexity of the probability distributions by some measure and clarifying the condition that the quantum kernel method offers any advantage is desired.

\subsubsection{Variance dependence}
\begin{figure}[htbp]
  \begin{minipage}[b]{0.32\hsize}
    \centering
    \includegraphics[keepaspectratio, scale=0.36]{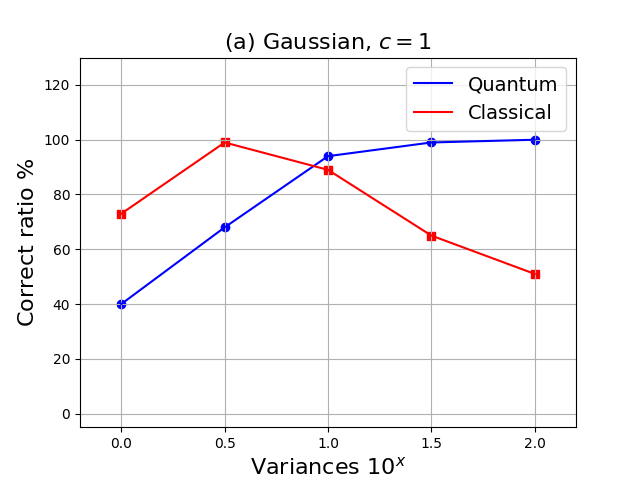}
    \label{figure:normal_accuracy_c_1}
  \end{minipage}
  \begin{minipage}[b]{0.32\hsize}
    \centering
    \includegraphics[keepaspectratio, scale=0.36]{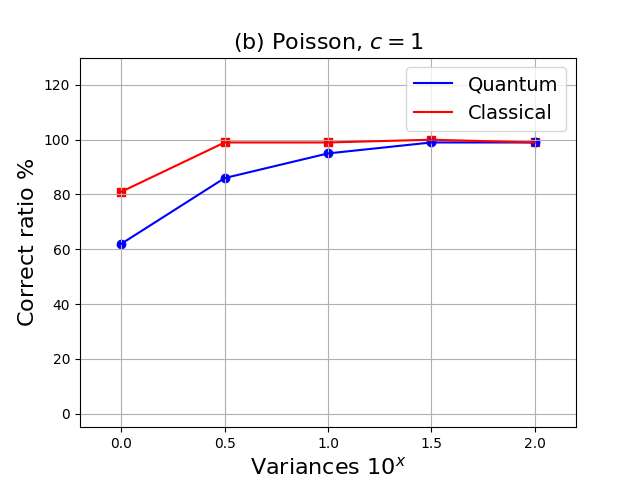}
        \label{figure:poisson_accuracy_c_1}
  \end{minipage}
  \begin{minipage}[b]{0.32\hsize}
    \centering
    \includegraphics[keepaspectratio, scale=0.36]{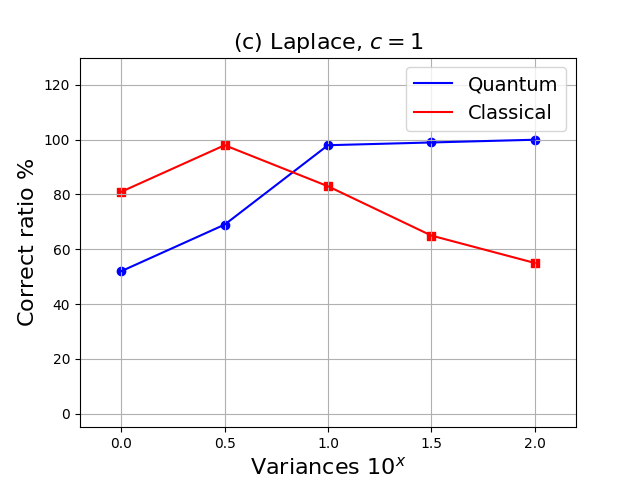}
    \label{figure:laplace_accuracy_c_1}
  \end{minipage}
  \label{figure:accuracy_diff_var_c_1}
  \begin{minipage}[b]{0.32\hsize}
    \centering
    \includegraphics[keepaspectratio, scale=0.36]{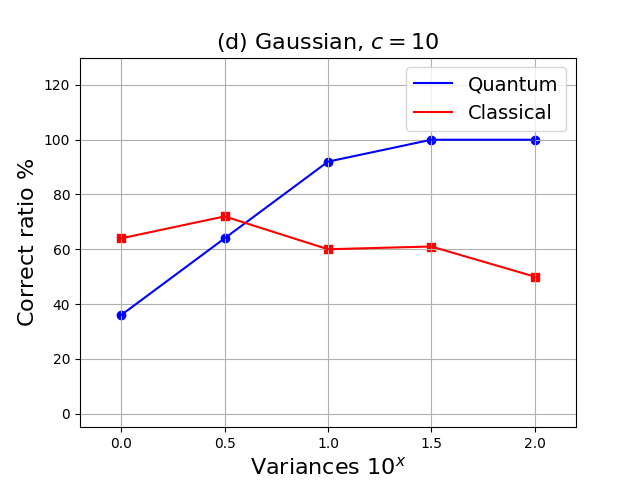}
    \label{figure:normal_accuracy_c_10}
  \end{minipage}
  \begin{minipage}[b]{0.32\hsize}
    \centering
    \includegraphics[keepaspectratio, scale=0.36]{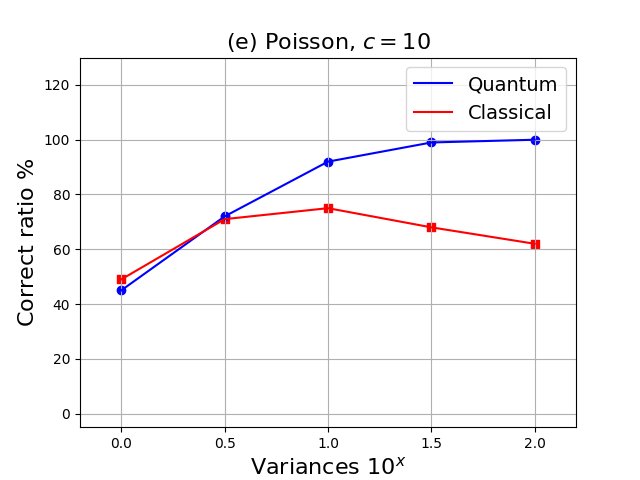}
        \label{figure:poisson_accuracy_c_10}
  \end{minipage}  
  \begin{minipage}[b]{0.32\hsize}
    \centering
    \includegraphics[keepaspectratio, scale=0.36]{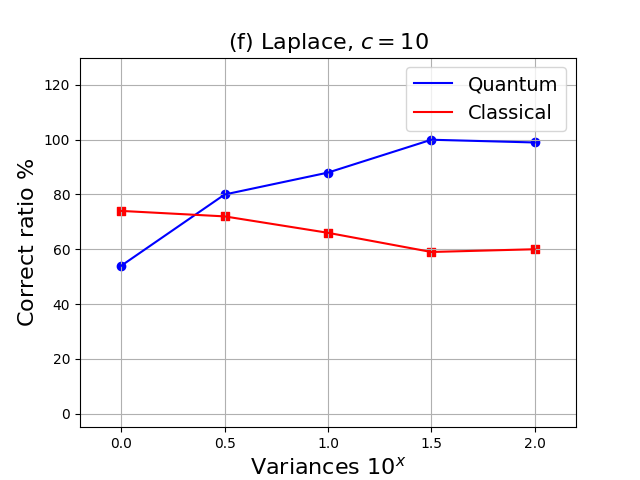}
    \label{figure:laplace_accuracy_c_10}
  \end{minipage}
  \begin{minipage}[b]{0.32\hsize}
    \centering
    \includegraphics[keepaspectratio, scale=0.36]{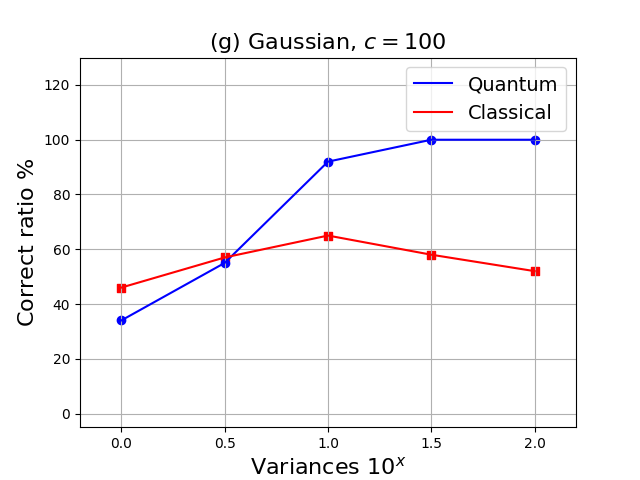}
    \label{figure:normal_accuracy}
  \end{minipage}
  \begin{minipage}[b]{0.32\hsize}
    \centering
    \includegraphics[keepaspectratio, scale=0.36]{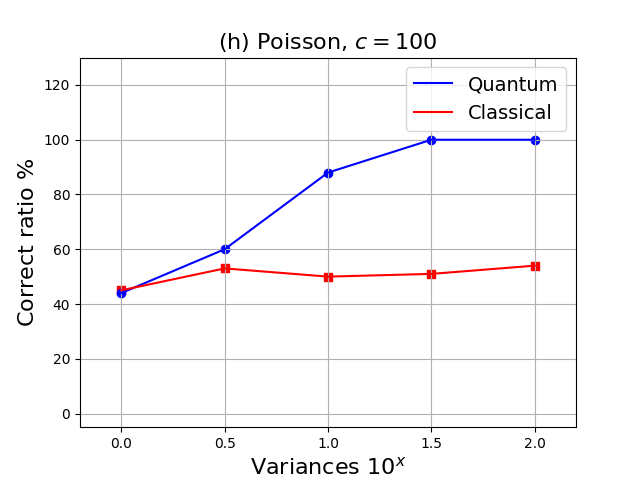}
        \label{figure:poisson_accuracy}
  \end{minipage}
  \begin{minipage}[b]{0.32\hsize}
    \centering
    \includegraphics[keepaspectratio, scale=0.36]{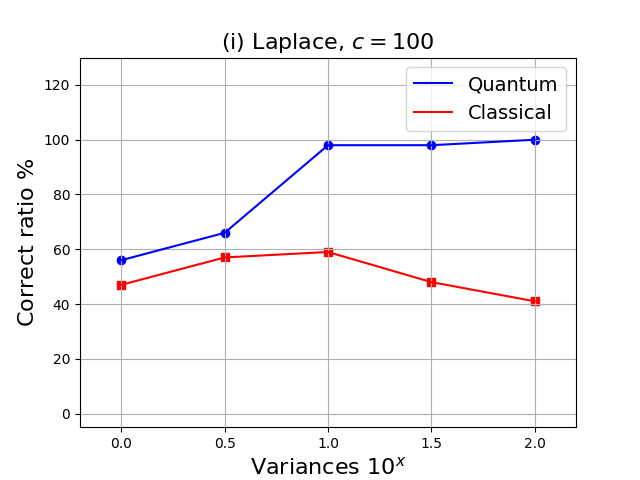}
    \label{figure:laplace_accuracy}
  \end{minipage}
  \caption{The correct ratio of the independence test with different variance. (Model: Linear, $N=10$)}
  \label{figure:accuracy_diff_var_c_10}
\end{figure}

\begin{figure}[htbp]
  \begin{minipage}[b]{0.32\hsize}
    \centering
    \includegraphics[keepaspectratio, scale=0.36]{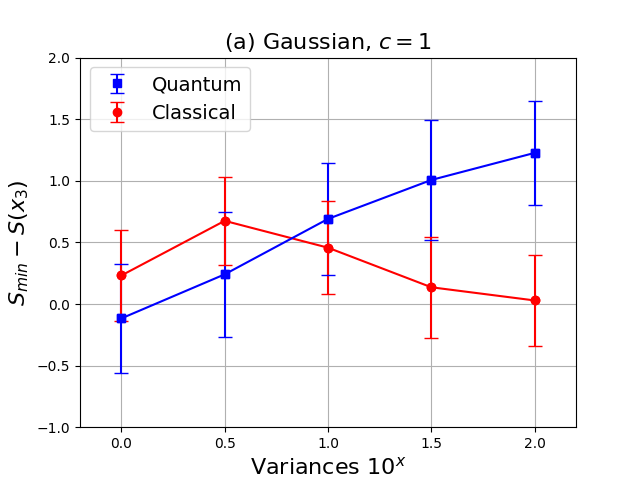}
    \label{figure:normal_mean_c_1}
  \end{minipage}
  \begin{minipage}[b]{0.32\hsize}
    \centering
    \includegraphics[keepaspectratio, scale=0.36]{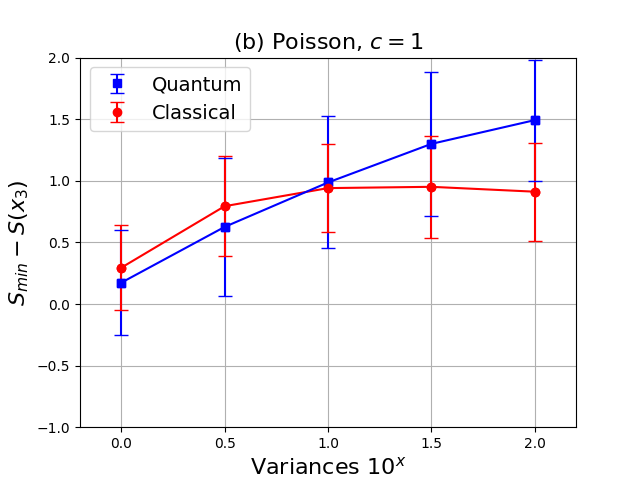}
    \label{figure:poisson_mean_c_1}
  \end{minipage}
  \begin{minipage}[b]{0.32\hsize}
    \centering
    \includegraphics[keepaspectratio, scale=0.36]{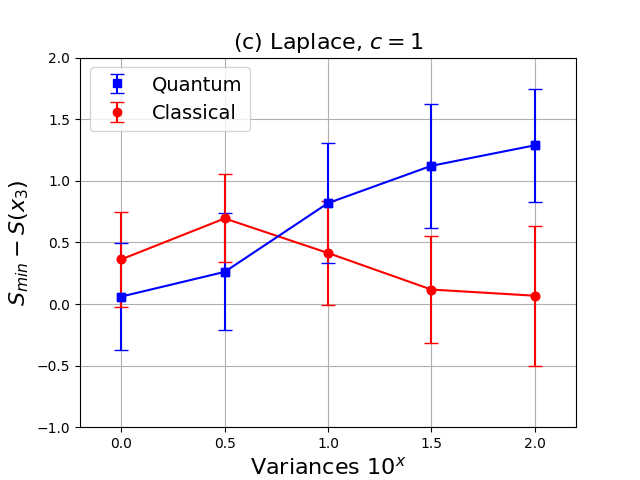}
    \label{figure:laplace_mean_c_1}
  \end{minipage}
  \begin{minipage}[b]{0.32\hsize}
    \centering
    \includegraphics[keepaspectratio, scale=0.36]{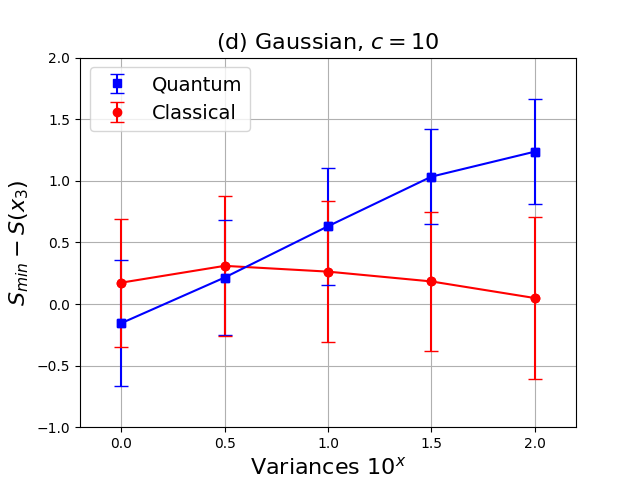}
    \label{figure:normal_mean_c_10}
  \end{minipage}
  \begin{minipage}[b]{0.32\hsize}
    \centering
    \includegraphics[keepaspectratio, scale=0.36]{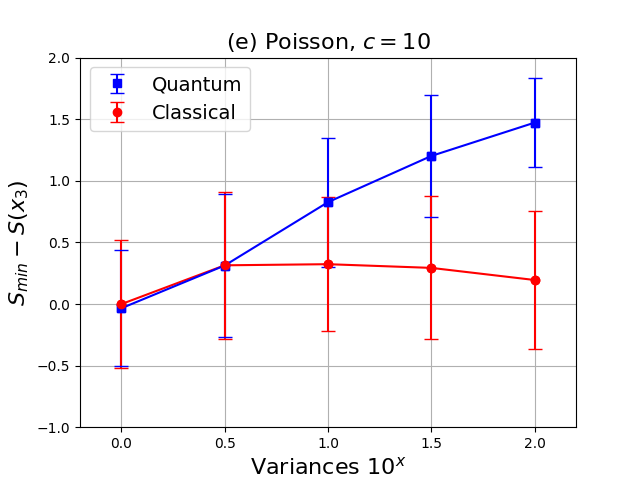}
    \label{figure:poisson_mean_c_10}
  \end{minipage}
  \begin{minipage}[b]{0.32\hsize}
    \centering
    \includegraphics[keepaspectratio, scale=0.36]{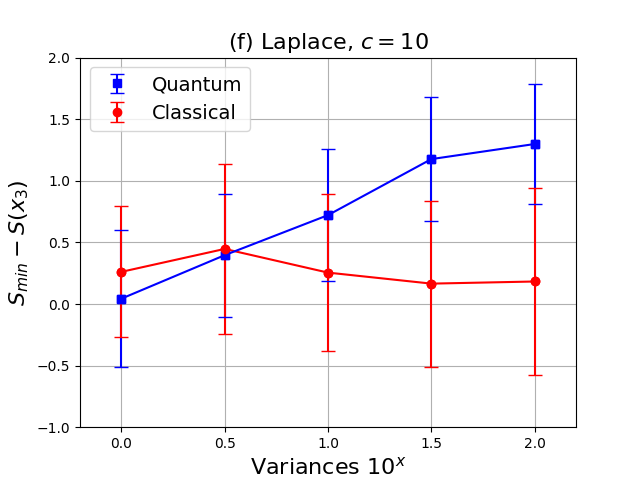}
    \label{figure:laplace_mean_c_10}
  \end{minipage}
  \begin{minipage}[b]{0.32\hsize}
    \centering
    \includegraphics[keepaspectratio, scale=0.36]{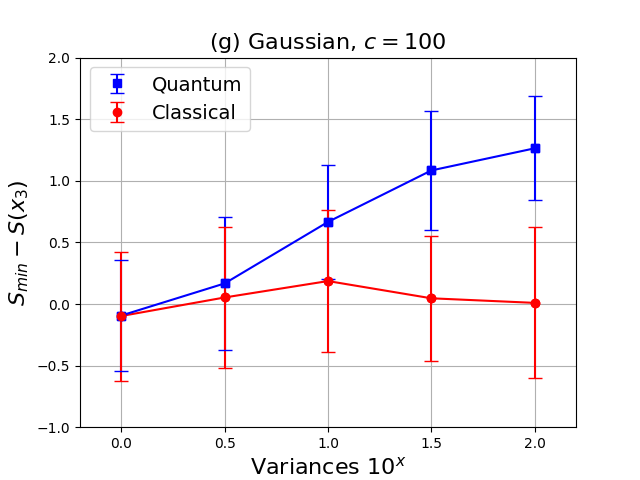}
    \label{figure:normal_mean}
  \end{minipage}
  \begin{minipage}[b]{0.32\hsize}
    \centering
    \includegraphics[keepaspectratio, scale=0.36]{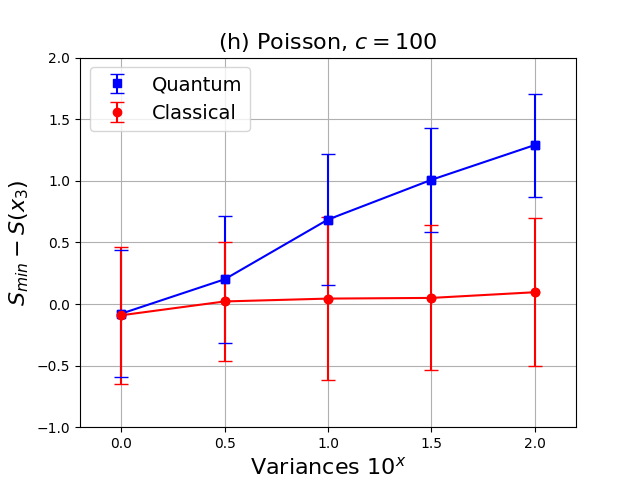}
    \label{figure:poisson_mean}
  \end{minipage}
  \begin{minipage}[b]{0.32\hsize}
    \centering
    \includegraphics[keepaspectratio, scale=0.36]{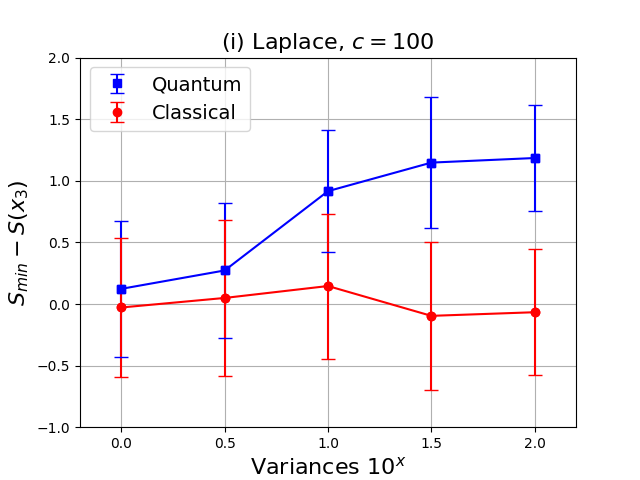}
    \label{figure:laplace_mean}
  \end{minipage}
    \caption{The slack values of the independence test with different variance. (Model: Linear, $N=10$)}
    \label{figure:slack_diff_var_c_10}
\end{figure}

FIG. \ref{figure:accuracy_diff_var_c_10} and \ref{figure:slack_diff_var_c_10} represent the variance dependence of correct ratio and slack values with the model function ``Linear", respectively.
It can be seen that the quantum kernel method performs better as the variance is increased under all of the condition, so that nature seems robust.
On the other hand, the performance of the classical kernel method degrades as the coefficient or the variance gets large.
These different behaviors would be related to the scope of the central limit theorem.
The theoretical background of the independence test with the classical kernel method relies on that theorem, but to meet the condition where it holds, a sufficient size of samples is necessary, and the specific sample size heavily depends on the variable distribution.
The quantitative and theoretical analysis from this light is also desired to clarify the property of the quantum kernel in this task.

\end{document}